\documentclass[12pt]{iopart}

\usepackage{graphicx}
\usepackage{amssymb}
\usepackage{amsfonts}
\usepackage{bm}
\usepackage{color}
\usepackage{multirow}
\usepackage{url}

\begin{document}

\title[Aerodynamic ground effect in fruitfly sized insect takeoff]{Aerodynamic ground effect in fruitfly sized insect takeoff}

\author{Dmitry Kolomenskiy$^{1,6}$\footnote{Corresponding
author: dkolom@gmail.com}, Masateru Maeda$^2$, Thomas Engels$^{3,4}$, Hao Liu$^{1,5}$, Kai Schneider$^3$ and Jean-Christophe Nave$^6$
\address{$^1$Graduate School of Engineering, Chiba University, 1-33 Yayoi-cho, Inage-ku, Chiba-shi, Chiba 263-8522, Japan}
\address{$^2$Department of Biology, Lund University, Ecology Building, SE-223 62 Lund, Sweden}
\address{$^3$M2P2--CNRS, Universit\'e d'Aix-Marseille, 39, rue Fr\'ed\'eric Joliot-Curie, 13453 Marseille Cedex 13, France}
\address{$^4$Institut f\"ur Str\"omungmechanik und Technische Akustik (ISTA), TU~Berlin, M\"uller-Breslau-Str. 8, 10623 Berlin, Germany}
\address{$^5$Shanghai-Jiao Tong University and Chiba University International Cooperative Research Center, 800 Dongchuan Road, Minhang District, Shanghai, China}
\address{$^6$Department of Mathematics and Statistics, McGill University, 805 Sherbrooke W., Montreal, QC, H3A 0B9, Canada}
}

\ead{dkolom@gmail.com}

\begin{abstract}
Aerodynamic ground effect in flapping-wing insect flight is of importance to comparative morphologies and of interest to the micro-air-vehicle (MAV) community. Recent studies, however, show apparently contradictory results of either some significant extra lift or power savings, or zero ground effect. Here we present a numerical study of fruitfly sized insect takeoff with a specific focus on the significance of leg thrust and wing kinematics.
Flapping-wing takeoff is studied using numerical modelling and high performance computing.
The aerodynamic forces are calculated using a three-dimensional Navier--Stokes solver
based on a pseudo-spectral method with volume penalization.
It is coupled with a flight dynamics solver that accounts for the body weight, inertia and the leg thrust,
while only having two degrees of freedom: the vertical and the longitudinal horizontal displacement.
The natural voluntary takeoff of a fruitfly is considered as reference.
The parameters of the model are then varied to explore possible effects of interaction between the flapping-wing model and the ground plane.
These modified takeoffs include cases with decreased leg thrust parameter, and/or with periodic wing kinematics, constant body pitch angle.
The results show that the ground effect during natural voluntary takeoff is negligible.
In the modified takeoffs,
when the rate of climb is slow, the difference in the aerodynamic forces
due to the interaction with the ground is up to 6\%.
Surprisingly, depending on the kinematics, the difference is either positive or negative,
in contrast to the intuition based on the helicopter theory, which suggests positive excess lift.
This effect is attributed to unsteady wing-wake interactions. A similar effect is found during hovering.
\end{abstract}


\maketitle

\section{Introduction}
The aerodynamic forces of an air vehicle or an animal may be affected by the ground proximity.
This phenomenon, known as the ground effect, has been extensively studied for aircraft \cite{Asselin_1997} and rotorcraft \cite{Leishman_2006}.
Although the effect varies depending on many design parameters, the general trend is an increase in lift and pitching moment, and a decrease in drag. The effect decays as the distance from the ground increases, and vanishes at a distance slightly larger than the characteristic length of the vehicle. For example, for a hovering helicopter, the excess thrust vanishes if the distance to the ground exceeds 1.25 times the diameter of the main rotor \cite{Leishman_2006}.

Rayner~\cite{Rayner_1991} proposed a fixed wing lifting line theory for forward flight of birds, bats and insects. His analysis suggested
that flight in ground effect provides performance improvements, if the flight speed is not too low.
However, this theory could not be applied to hovering or slow forward flight at very low height, since it neglected flapping motion.
Normal hovering in ground effect was considered by Gao and Lu \cite{Gao_Lu_2008}.
They carried out two-dimensional numerical simulations of hovering and
identified three regimes: force enhancement,
force reduction, and force recovery, depending on the distance from the ground.
Liu \emph{et al.} \cite{Liu_etal_2009} considered clap-and-fling near the ground and found force enhancement at all distances.
A three-dimensional numerical simulation of fruitfly hovering was carried out by Maeda and Liu \cite{Maeda_Liu_2013}.
An increase in lift and a reduction in power was found.
A significant vertical force was generated on the insect's body due to the `fountain effect'.
Energetic savings have also been reported for a hummingbird hovering in ground effect \cite{Kim_etal_2015}.

Several studies considered pitching-plunging foils near a solid wall or a free surface \cite{Tanida_2001,Quinn_etal_2014,Wu_etal_2014}.
This configuration is relevant to fish swimming as well as forward flapping flight.
The ground effect mainly consists in enhanced propulsive force.
However, it also generates a non-zero vertical force due to asymmetry.

The main motivation for this study comes from the fact that the ground proximity is natural for takeoff and landing.
These manoeuvres, unlike hovering or forward flight, are characterized by gradual change of distance to the ground.
The `dynamic' ground effect in these circumstances may be different from the `static' effect at a constant distance \cite{Jones_PhD_2005}.
This difference may be even larger for flapping wings than for fixed wings, because animals vary their wing kinematics during takeoff.

So far, the ground effect during takeoff has been assessed for very few insects only.
It was found negligible for butterflies (\textit{Pieris rapae} \cite{Bimbard_etal_2013}, \textit{Papilio xuthus} \cite{Maeda_PhD_2014}), a dronefly (\textit{Eristarlis tenax}) \cite{Chen_etal_2013}, and a fruitfly (\textit{Drosophila virilis}) \cite{Chen_Sun_2014}, but significant for a beetle (\textit{Trypoxylus dichotomus}) \cite{Truong_etal_2013}.
The disparity can be attributed to significant differences in the size, morphology and kinematics of these insects.
Thus, our work is motivated by the apparently contradictory conclusions
on the significance of the ground effect that could be found in
the animal flight literature. It is important to identify the parameters that make the ground effect strong or negligible.

In the present study, we consider a numerical model having the morphology of a fruitfly, with variable wing kinematics and leg parameters.
Our objective is to determine if the ground effect can be significant for this model, and which conditions can lead to it.
We thus explore the parameter space of the model and perform numerous numerical simulations using \textsc{FluSi} \cite{Engels_etal_2015},
which is an open source software available on \url{https://github.com/pseudospectators/FLUSI/tree/plos_one_ground_effect}.
First, for completeness, we revisit the voluntary takeoff of a fruitfly analyzed in \cite{Chen_Sun_2014}.
The main difference with respect to \cite{Chen_Sun_2014} is the use of a flight dynamics solver.
We then compare takeoffs with modified parameters of the leg thrust model and wing kinematics.
Finally, we consider hovering as a limiting case of very slow takeoff.

The paper is organized as follows.
In section~\ref{sec:models_and_methods}, we describe our computational approach and the takeoff parameters used in this study.
The results are presented in section~\ref{sec:results}, first for a natural voluntary takeoff, then for modified takeoffs and for hovering flight.
The main conclusions are summarized in section~\ref{sec:conclusions}.

\section{Methods}\label{sec:models_and_methods}

\subsection{Morphology and kinematics}

\begin{figure}
\begin{center}
\includegraphics{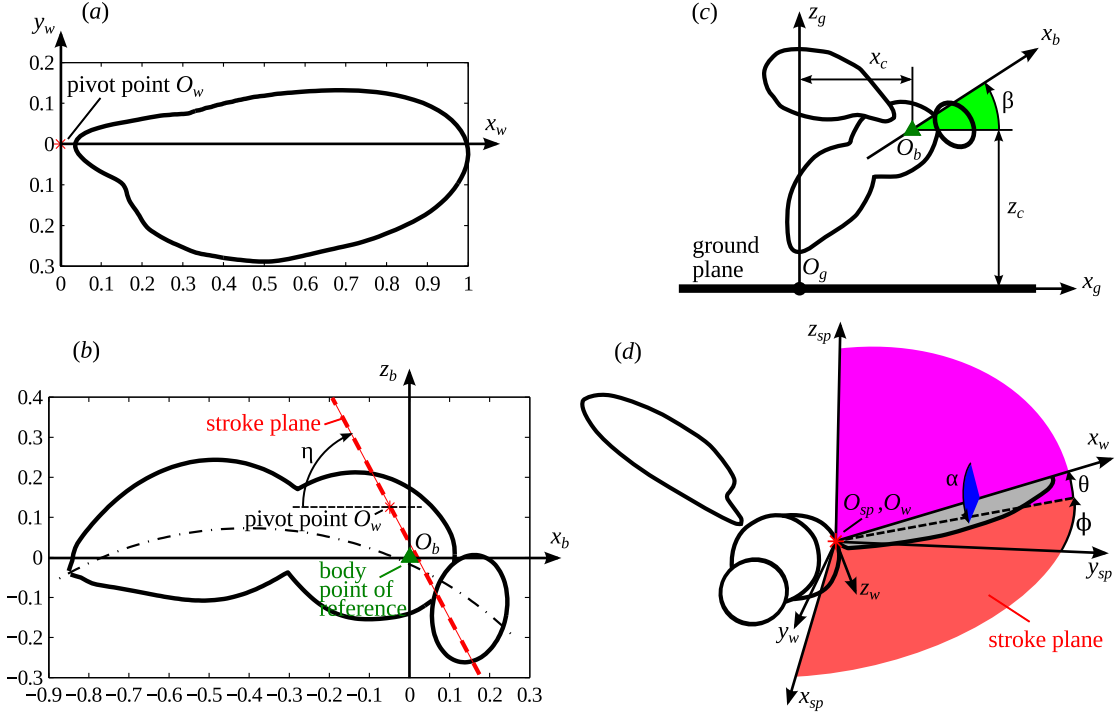}
\caption{\label{fig:kinangles} {\bf Schematic drawing of the morphological model.} (\textit{a}) Wing contour. Coordinates are normalized to the wing length $R$.
(\textit{b}) The body is generated by circular sections of variable radius, which changes depending on the position along the centre line (dash-dotted arc).
The body axis $O_b x_b$ is the thorax-abdomen principal axis, approximately.
The coordinates of the wing pivot points in the body frame of reference $O_{b}x_{b}y_{b}z_{b}$ are $(-0.07R,\pm0.18R,0.115R)$.
Note that, since $\beta(t)$ is prescribed in the simulations and only $(x_c,z_c)$ are dynamically calculated,
the position of the body point of reference with respect to the body contour is chosen arbitrarily.
(\textit{c}) The insect's position
with respect to the ground is described by the body point of reference coordinates $(x_c,z_c)$ and the position angle $\beta$.
(\textit{d}) Definition of the wing's angles with respect to the stroke plane frame of reference $O_{sp}x_{sp}y_{st}z_{sp}$. The origin $O_{sp}$ is the wing pivot point.}
\end{center}
\end{figure}

In this work, we consider a fruitfly having mass $m=1.2$ mg and wing length $R=2.83$ mm,
which are the values reported by Chen and Sun \cite{Chen_Sun_2014}.
The body is modelled as a rigid solid, and the wings are modelled as rigid flat plates.
This approximation is accurate for drosophila during voluntary takeoff \cite{Fontaine_etal_2009,Chen_Sun_2014},
though it occasionally fails during fast escape manoeuvres \cite{Fontaine_etal_2009}.
The wing contour used in this study is shown in Fig.~\ref{fig:kinangles}(\textit{a}). It is adapted from \cite{Chen_Sun_2014}. Its mean chord length is equal to $c=0.85$ mm.
The body is generated by sweeping a circular section of variable radius along a curvilinear centreline (an arc).
The body has approximately the same dimensions as in \cite{Chen_Sun_2014}.
The side view of the body is shown in Fig.~\ref{fig:kinangles}(\textit{b}).
Even though the yaw and roll angles can eventually become large during takeoff, there is no significant
trend for all takeoffs. Hence, to simplify the problem, we assume bilateral symmetry.
Therefore, the body orientation is fully defined by the pitch angle $\beta$ between the body and the horizontal axis, see Fig.~\ref{fig:kinangles}(\textit{c}).
The wing kinematics is described by three angles: $\phi$, $\alpha$ and $\theta$, measured with respect to the stroke plane, as shown in Fig.~\ref{fig:kinangles}(\textit{d}).
The positional angle $\phi$ defines the motion of the wing tip projection on the stroke plane.
The deviation (elevation) angle $\theta$ defines the deviation of the wing tip from the stroke plane.
The feathering angle $\alpha$ defines the rotation about the longitudinal axis of the wing, and it is related to the geometrical
angle of attack ($AoA$) as $\alpha=90^\circ-AoA$ during downstroke and as $\alpha=90^\circ+AoA$ during upstroke.
It is convenient to refer to an `anatomical' stroke plane angle $\eta$, i.e.,
to assume that the inclination of the stroke plane against the body axis is held at a constant angle for any motion of the body. 

Since the main focus of this study is the ground effect, it is important to ensure
that the time evolution of the distance to the ground is consistent with the forces acting on the insect.
For this reason, in our computations, unlike in \cite{Chen_Sun_2014}, the position of the insect is dynamically computed as opposed to be prescribed.
We compute the position of the body point of reference $(x_c,z_c)$, see Fig.~\ref{fig:kinangles}(\textit{b}), from Newton's 2nd Law,
\begin{equation}
m \frac{\mathrm{d}^2 x_c}{\mathrm{d} t^2} = F_{ax} + F_{\ell x}, \quad\quad m \frac{\mathrm{d}^2 z_c}{\mathrm{d} t^2} = F_{az} + F_{\ell z} - mg,
\label{eq:newton}
\end{equation}
where $(F_{ax},F_{az})$ is the aerodynamic force, $(F_{\ell x},F_{\ell z})$ is the leg thrust, subscripts $x$ and $z$ correspond, respectively, to the horizontal and vertical components,
$m$ is the insect's mass and $g$ is the gravitational acceleration. Equations (\ref{eq:newton}) are integrated using the adaptive second order Adams--Bashforth scheme \cite{Schneider_2005}, simultaneously with the incompressible Navier--Stokes equations.
We defined the positive $z$ direction to be upwards and the positive $x$ direction to be forwards (see figure~\ref{fig:kinangles}\textit{c}).

\subsection{Aerodynamics}

The aerodynamic forces $F_{ax}$ and $F_{az}$ are obtained by solving the three-dimensional incompressible Navier--Stokes equations.
The no-slip boundary condition at the body and wings surfaces is imposed using the volume penalization method \cite{Angot_etal_1999},
and the penalized equations are solved using a classical Fourier pseudo-spectral method.
More details about the solver and the generic insect model, including a numerical validation case of fruitfly hovering, can be found in \cite{Engels_etal_2015}.
Numerical validation of the ground plane modelling using the volume penalization method is described in \ref{sec:validation}.

The computational domain in the present study is a rectangular box with sides $L_x$, $L_y$ and $L_z$.
Suitable values of $L_x$, $L_y$ and $L_z$, in terms of accuracy and computational efficiency, depend on the motion of the insect within the domain. Therefore, different values
are used in different simulations, as described later in the text.
The domain is discretized using a uniform Cartesian grid.
Periodic boundary conditions are applied on all sides of the domain, as
required by the Fourier discretization.
Vorticity sponge boundary conditions are imposed at the left, right, rear and front sides of the domain, as explained in \cite{Engels_etal_2014,Engels_etal_2015},
in order to minimize the effect of the finite domain size.
The ground surface is modelled as a solid layer at the bottom of the domain,
which by periodicity also imposes the no-slip on the top of the fluid domain.
We have carried out numerical experiments to ensure that, in the numerical simulations presented in this paper, the domain size is sufficiently large, i.e., its further increase does not change significally the forces. The dimensions that we chose are also comparable with the size of the mineral oil tank
used in the experiments with a mechanical model \cite{Dickinson_etal_1999}.

\subsection{Leg thrust}

The model of the leg thrust employed in the present study is a slight modification of the compression spring model proposed in \cite{Bimbard_etal_2013}.
We assume that takeoff begins from rest and starts at time $t=t_\ell$, which can be estimated from the initiation of the legs motion in the video sequences shown in \cite{Chen_Sun_2014}.
The two components of the force are given by
\begin{equation}
F_{\ell x} = Z \cot \phi_\ell, \quad\quad F_{\ell z} = Z.
\label{eq:legs_thrust}
\end{equation}
The magnitude of the leg force is assumed to depend on the vertical component of the leg extension $\zeta = z_c(t)-z_c(t_\ell)$ only.
The force is supposed to be distributed between the three pairs of legs such that its change with horizontal displacement can be neglected,
\begin{equation}
Z = \left\{
\begin{array}{ll}
\displaystyle (L_\ell - \zeta) K_\ell & \textrm{for  } \zeta < L_{\ell}, \\
\displaystyle 0 & \textrm{for  } \zeta \ge L_{\ell}, \\
\end{array}
\right.
\label{eq:legs_thrust_magnitude}
\end{equation}
where $L_\ell$ is the maximum leg extension length, i.e., the difference between the values of $z_c$ when the legs are fully extended at takeoff
and when the insect is at rest. When the legs are fully extended, $\zeta = L_\ell$, the legs lose contact with the ground and the force drops to zero.
This length is estimated using video sequences in \cite{Chen_Sun_2014} to be equal to $L_\ell=1.24$~mm.
The spring stiffness $K_\ell$ varies in time: it increases from $K_\ell^-$ before takeoff to $K_\ell^+$ after takeoff.
The initial value $K_\ell^-=mg/L_\ell$ ensures that the insect is in equilibrium before takeoff, when the aerodynamic force is zero.
The final value $K_\ell^+$ is a parameter of the model that controls the maximum leg thrust.
Its value can be estimated from the climb velocity at the beginning of takeoff, shown in, e.g., \cite{Chen_Sun_2014}.
It may also be estimated from jumps of wingless flies \cite{Zumstein_etal_2004,Card_Dickinson_2008} for a slightly different fruitfly, \textit{D. melanogaster}.
We assume the time evolution of $K_\ell$ of the form
\begin{equation}
K_{\ell} = \left\{
\begin{array}{ll}
\displaystyle K_{\ell}^- & \textrm{for  } t < t_{\ell}, \\
\displaystyle K_{\ell}^- + \frac{K_{\ell}^+-K_{\ell}^-}{\tau_\ell} (t-t_{\ell}) & \textrm{for  } t_{\ell} \le t < t_{\ell}+\tau_{\ell}, \\
\displaystyle K_{\ell}^+ & \textrm{for  } t \ge t_{\ell}+\tau_{\ell}.
\end{array}
\right.
\label{eq:legs_thrust_K}
\end{equation}
The transition time $\tau_{\ell}$ can be equal to zero, in which case the leg force increases impulsively at the beginning of takeoff.
However, measurements of the leg force \cite{Zumstein_etal_2004} suggest a gradual increase which can be accounted for by setting $\tau_\ell$ to a value larger than zero.
The value $\tau_\ell=1.3$ms results in the gradient $\mathrm{d}F_{\ell}/\mathrm{d}t$ consistent with the experimental data shown in \cite{Zumstein_etal_2004}.
The direction $\phi_\ell$ also changes in time. Before takeoff, when the insect is at rest, the force is applied only in the vertical direction, i.e., $\phi_\ell^- = 90^\circ$.
During takeoff, the horizontal component is non-zero, in general. 
We assume a time evolution of the form
\begin{equation}
\phi_{\ell} = \left\{
\begin{array}{ll}
\displaystyle \phi_{\ell}^- & \textrm{for  } t < t_{\ell}, \\
\displaystyle \phi_{\ell}^- + \frac{\phi_{\ell}^+-\phi_{\ell}^-}{\tau_\ell} (t-t_{\ell}) & \textrm{for  } t_{\ell} \le t < t_{\ell}+\tau_{\ell}, \\
\displaystyle \phi_{\ell}^+ & \textrm{for  } t \ge t_{\ell}+\tau_{\ell}.
\end{array}
\right.
\label{eq:legs_thrust_phi}
\end{equation}
The values of the leg thrust model parameters used in our numerical simulations are given in table~\ref{tab:legs_params}.


\section{Results and Discussion}\label{sec:results}

\subsection{Summary of the numerical simulations}\label{sec:summary_simulations}

The starting point for our study is the voluntary takeoff, as it is shown in section~\ref{sec:voluntary_takeoff} (in agreement with \cite{Chen_Sun_2014}) that the ground effect is very small in that case.
It is much smaller than during hovering (cf. \cite{Maeda_Liu_2013}).
We conjecture that this difference is due to the large takeoff vertical velocity, which is mainly the result of the leg thrust.
To test this hypothesis, in section~\ref{sec:slow_vertical_takeoff}, we discuss a situation in which the legs produce less force and the insect takes off slower.
The ground effect becomes significant.
The vertical force increases during the first two wingbeats due to the ground effect,
but slightly decreases later on.
We then carry out a parametric study using periodic wing kinematics in section~\ref{sec:simple},
and find an even stronger adverse ground effect.
Finally, in section~\ref{sec:hovering}, we find similar trends during the first wingbeats in hovering flight,
which can is considered as a limiting case of takeoff with zero rate of climb.
Table~\ref{tab:legs_params} summarizes the parameters of the different cases considered in the present study.
Datasets for the `voluntary' and `simplified' cases can be downloaded from ResearchGate at \url{http://dx.doi.org//10.13140/RG.2.1.2145.2562}.

\begin{table}[h]
\caption{\bf{ Parameters of the takeoffs considered in the present study.}}
\begin{tabular}{|l|l|l|l|l|l|l|l|l|l|}
\hline
{\bf Case}            & \multicolumn{4}{l|}{{\bf Kinematics}}      & \multicolumn{5}{l|}{{\bf Leg model}} \\ \hline 
Name                  & $\phi$, $\alpha$, $\theta$           & $\beta$                            & $\eta$,$^\circ$ & $z_c(0)$, mm & $L_\ell$, mm & $K_\ell^+$, N/m    & $\phi_\ell^+$,$^\circ$ & $t_\ell$, ms & $\tau_\ell$, ms \\ \hline
Voluntary             & fig.\ref{fig:kinangles_upd}          & fig.\ref{fig:kinangles_upd}        & 62               & 1.08         & $1.24$       & $0.165$            & $84$                    & $4.2$        & $1.3$           \\ \hline
Slow                  & fig.\ref{fig:kinangles_upd}          & fig.\ref{fig:kinangles_upd}        & 62               & 1.08         & $1.24$       & $0.041$            & $84$                    & $4.2$        & $1.3$           \\ \hline
Simplified            & fig.\ref{fig:kine}(\textit{a})       & $46.3^\circ$                       & 32               & 3.11         & $1.24$       & $0.0095$...$0.043$ & $84$                    & $0$          & $1.3$           \\ \hline
Hovering              & fig.\ref{fig:kine}(\textit{a,b})     & $55^\circ$                         & 55               & 2.07         & N/A          & N/A                & N/A                     & N/A          & N/A             \\ \hline
\end{tabular}

\begin{tabular}{|l|l|l|l|l|l|l|l|l|l|}
\hline
{\bf Case}            & \multicolumn{7}{l|}{{\bf Numerical parameters}} \\ \hline 
Name                  & $L_x$ & $L_y$ & $L_z$ & $N_x$ & $N_y$ & $N_z$  & $\varepsilon$       \\ \hline
Voluntary             & $5R$  & $5R$  & $8R$  & $640$ & $640$ & $1280$ & $2.5 \cdot 10^{-4}$ \\ \hline
Slow                  & $5R$  & $5R$  & $6R$  & $640$ & $640$ & $768$  & $2.5 \cdot 10^{-4}$ \\ \hline
Simplified            & $4R$  & $4R$  & $6R$  & $512$ & $512$ & $768$  & $2.5 \cdot 10^{-4}$ \\ \hline
Hovering              & $8R$  & $8R$  & $4R$  & $864$ & $864$ & $432$  & $2.5 \cdot 10^{-4}$ \\ \hline
\end{tabular}
\label{tab:legs_params}
\end{table}

\subsection{Voluntary takeoff}\label{sec:voluntary_takeoff}

In this section, we consider voluntary takeoff of a fruitfly with the parameters as in the first line in Table~\ref{tab:legs_params}.
This case shows some important general features of fruitfly takeoff
such as the first wingbeat cycles beginning while the legs extend.
Therefore it is likely that, despite some variability in voluntary takeoffs,
the ground effect in general
remains of the same order of magnitude in natural circumstances.

\begin{figure}
\begin{center}
\includegraphics{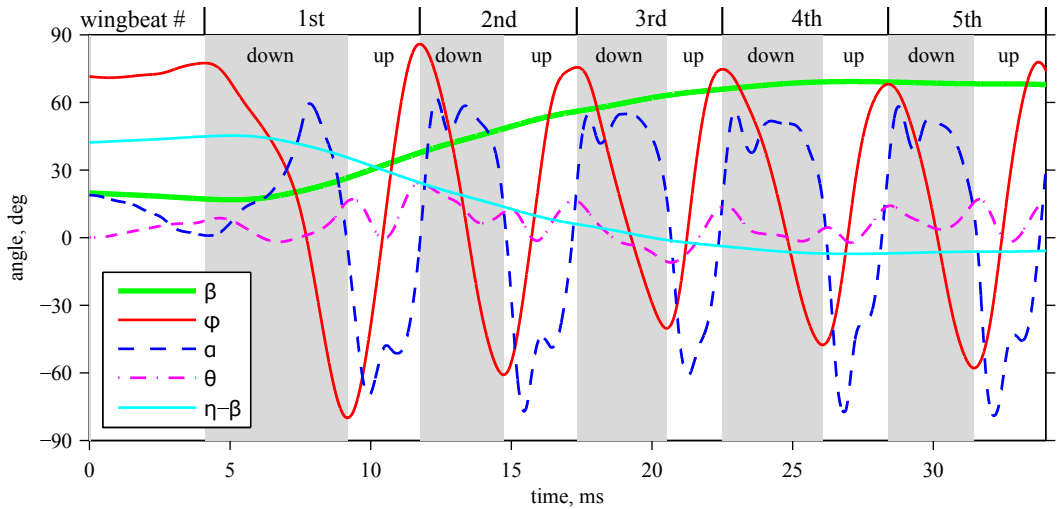}
\caption{\label{fig:kinangles_upd}{\bf Time evolution of the angular position of the body and of the wings during the voluntary takeoff.} Gray shaded regions correspond to downstrokes. $\eta-\beta$ is the angle
between the horizontal plane and the stroke plane, i.e., the global stroke plane angle \cite{Tobalske_etal_2007}.}
\end{center}
\end{figure}

The values of the body and wing angles are taken from one of the
cases documented in \cite{Chen_Sun_2014}.
However, the wing motion in \cite{Chen_Sun_2014} is not exactly symmetric.
Therefore, the time series of $\phi$, $\alpha$ and $\theta$ that we use for both wings correspond to the left wing data shown in \cite{Chen_Sun_2014}.
Fig.~\ref{fig:kinangles_upd} presents the time evolution of the wing positional angle $\phi(t)$,
the feathering angle $\alpha(t)$, the elevation angle $\theta(t)$ and the body pitch angle $\beta(t)$, which are prescribed in our numerical simulations. The angle between the horizontal plane and the stroke plane, $\eta-\beta$,
is also shown for reference.

Even though the wing motion is not exactly periodic, it is useful to introduce the wing beat frequency.
When calculated using the average wing beat cycle period over the five cycles shown in Fig.~\ref{fig:kinangles_upd},
it is equal to $f=169$~Hz. Similarly, the average wing beat amplitude is equal to $\Phi=134^\circ$, and the characteristic wing tip velocity is $U=2 \Phi R f=2.23$~m/s.
The kinematic viscosity of air, equal to $\nu=1.45\cdot10^{-5}~\textrm{m}^2/\textrm{s}$ yields the Reynolds number $Re=Uc/\nu=131$.
Note that $U$ and $Re$ do not account for the forward speed of the body.

The computational domain size is equal to $L_x=L_y=5R$, $L_z=8R$, where $R$ is the wing length. 
The influence of the domain size in the vertical direction is discussed in \ref{sec:domain_size}.
The number of grid points in each direction, respectively, is $N_x=N_y=640$ and $N_z=1280$. The penalization parameter is $\varepsilon = 2.5\cdot10^{-4}$ (for details see, e.g., \cite{Engels_etal_2014}).

The aerodynamic ground effect is evaluated by comparing two numerical simulations with two different values of the initial distance from the body point of reference to the ground:
$z_c(0)=0.38R$ and $2R$, which we denote `in ground effect' (IGE) and `out of ground effect' (OGE), respectively. The first case corresponds to a takeoff from a flat ground surface,
with $z_c(0)$ being consistent with the data in \cite{Chen_Sun_2014}. In the second case, the leg model behaves as during takeoff from the ground,
but the aerodynamic interaction between the insect and the ground is weak because of the large distance. This case may be interpreted as takeoff from a perch
that provides enough support for the legs but has a small surface, such that the aerodynamic interactions are negligible.
With the distance equal to $2R$ or greater, the ground effect is negligible during hovering \cite{Maeda_PhD_2014}.
The circulation of the wake vortices is mainly determined by the integral aerodynamic force, therefore it is not larger during takeoff than during hovering,
and the spatial rate of decay of the induced velocity is the same. Hence, the ground effect with the distance equal to $2R$ is likely to be negligible during takeoff.
The influence of the ground on the shape of the vortices is only visible during the 2nd wingbeat and later on.
This difference is localized to the vicinity of the ground plane. Since the insect is relatively far from the ground by that time,
this difference is unlikely to have any influence on the aerodynamic forces.

\begin{figure}
\begin{center}
\includegraphics[scale=0.85]{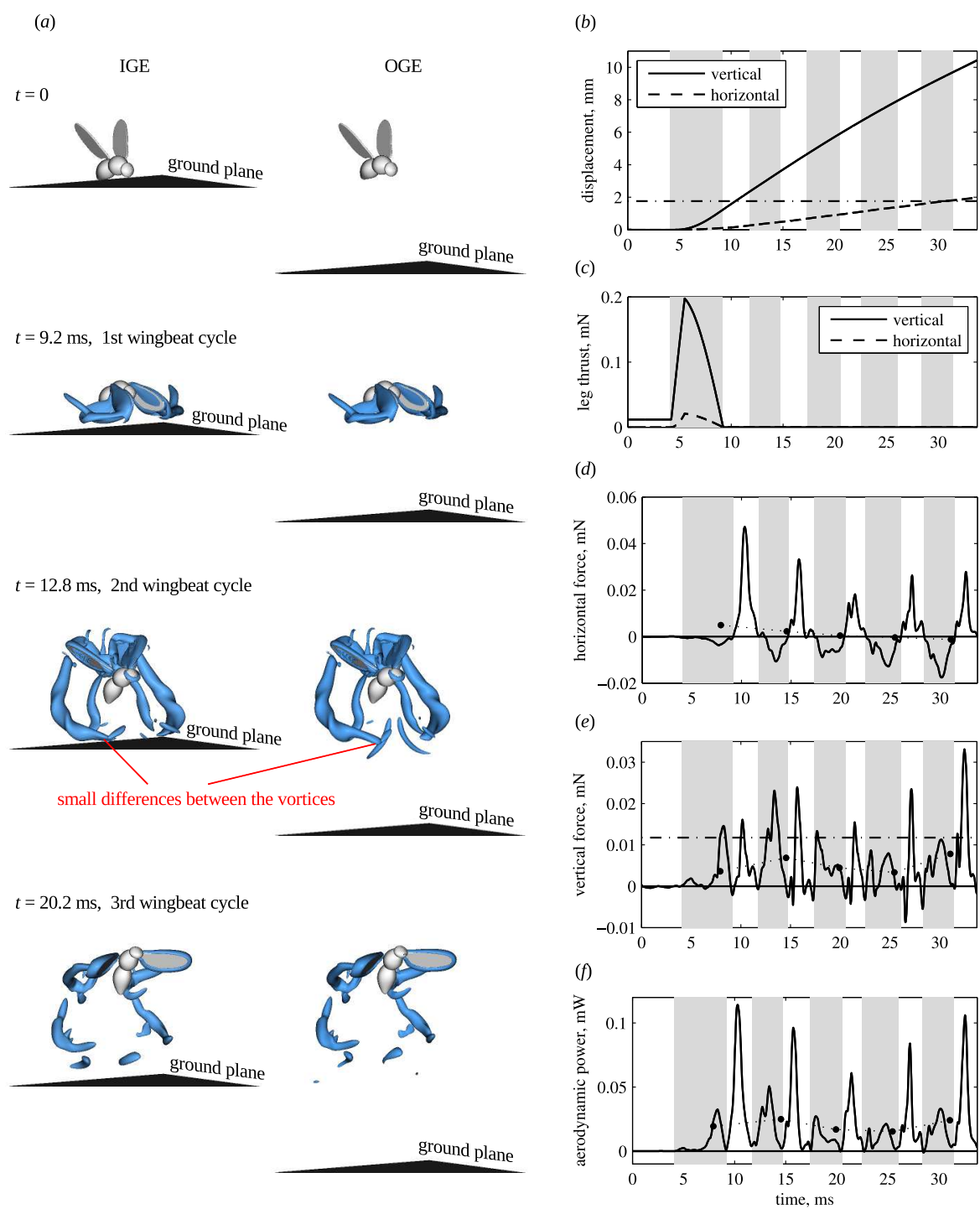}
\caption{\label{fig:drawing_voluntary} {\bf Voluntary takeoff.} (\textit{a}) Visualization of the wings, body and ground surface, and the wake at 4 subsequent time instants.
Blue semi-transparent iso-surfaces show the $Q$-criterion, $Q/f^2=15$. (\textit{b}) Vertical and horizontal displacement. To obtain distance $z_c$ from the ground for the IGE case, add 1.08 mm. The black dash-dotted line indicates $z_c=R$. (\textit{c}) Components of the leg force. (\textit{d}) horizontal and (\textit{e}) vertical components of the aerodynamic force and (\textit{f}) the aerodynamic power. The black dash-dotted line in Fig. (\textit{e}) indicates the weight. Solid circles connected by dotted lines show wingbeat cycle averages. The results for OGE and IGE are shown, but the curves in Fig. (\textit{b-f}) overlap because the difference is negligible.} 
\end{center}
\end{figure}

Fig.~\ref{fig:drawing_voluntary}(\textit{a}) shows the fruitfly model and the wake, IGE and OGE, at 4 subsequent time instants.
The vortices created by the wings and the body are identified as the volume of fluid enclosed by the iso-surfaces of the $Q$-criterion. 
At $t=0$, the air is at rest. The insect body is almost horizontal. The wings are in a pre-takeoff position from which they begin the first downstroke after $t=4.1$~ms.
The time $t=9.2$~ms corresponds to the first reversal from downstroke to upstroke.
Because of the small body pitch angle $\beta$, the stroke plane is effectively vertical.
In addition, the wing tip speed during the first downstroke is smaller than during all subsequent strokes.
Therefore, the vertical aerodynamic force is small, but the body lifts noticeably because of the leg thrust.
The time $t=12.8$~ms corresponds to the second upstroke.
At this point, the distance from the body point of reference to the ground $z_c$ is already larger than the wing length $R$.
Therefore, the aerodynamic interference with the ground is expected to be very small.
Note that the kinematics during the first two wingbeat cycles are a transient. After that, the time evolution of the wing angles approaches a periodic regime and
the stroke plane becomes less inclined with respect to the ground, see Fig.~\ref{fig:kinangles_upd}. 

The displacement of the body point of reference is shown in Fig.~\ref{fig:drawing_voluntary}(\textit{b}).
It presents the evolution of the vertical component $\zeta(t)=z_c(t)-z_c(0)$ and the
horizontal component $\xi(t)=x_c(t)-x_c(0)$ over time for the cases IGE and OGE.
The curves overlap. In both cases, at the end of the 4th wingbeat cycle, $t=28.4$~ms, the insect gains 8.7~mm of altitude and propels 1.6~mm forward.
These numbers are consistent with the trajectories shown in \cite{Chen_Sun_2014}.
The displacements IGE and OGE differ by less than 1\%.
Therefore, the ground effect on $\zeta$ and $\xi$ is indeed negligible.

Fig.~\ref{fig:drawing_voluntary}(\textit{c}) shows the two components of the leg force.
At $t=0$, the vertical component of the leg force is equal to the weight and
the horizontal component is zero.
The jump is triggered at $t_\ell=4.2$~ms. At time $t_\ell+\tau_\ell=5.5$~ms, both components reach their peaks.
After that the force decreases and vanishes at $t=9.3$~ms, when the legs lose contact with the ground.
Note that the leg thrust can, in principle, be different for the takeoffs IGE and OGE, because the leg model depends on the aerodynamic force \textit{via} $z_c(t)$.
However, for the voluntary takeoff considered here, there is no influence of the ground effect.

The vertical and the horizontal components of the aerodynamic force are shown in Fig.~\ref{fig:drawing_voluntary}(\textit{d}) and (\textit{e}), respectively.
Over the first four wingbeat cycles, the wingbeat averaged aerodynamic forces are significantly lower than the weight.
This can be explained by the large initial rate of climb due to the leg thrust, which cannot be supported by the wings.
Even during the fourth wingbeat, the wing force is equal to 29\% of the weight. The vertical acceleration is therefore negative after the legs lose contact with the ground,
and the rate of climb slowly decreases.
The ground effect is, again, negligible. Even during the first wingbeat cycle, when the wings approach the ground surface, the difference in the instantaneous vertical force between IGE and OGE is at most 0.0005~mN, i.e., about 4\% of the weight.
The wingbeat cycle averaged forces differ by less than 1\% of the weight. 

Fig.~\ref{fig:drawing_voluntary}(\textit{f}) displays the time evolution of the aerodynamic power, when operating IGE and OGE.
Note that, in this study, we do not consider the inertial power because the wings have the same kinematics in both cases, IGE and OGE.
Therefore, the inertial power is the same.
The aerodynamic power is the aerodynamic component of the power required to actuate the wings,
\begin{equation}
P = - \bm{M}_l \cdot (\bm{\Omega}_l-\bm{\Omega}_b) - \bm{M}_r \cdot (\bm{\Omega}_r-\bm{\Omega}_b).
\label{eq:aero_power}
\end{equation}
In (\ref{eq:aero_power}), $M_l$ and $M_r$ are the aerodynamic moments of the left and of the right wing, respectively, relative to the corresponding pivot point.
$\Omega_l$ and $\Omega_r$ are the angular velocities of the wings and $\Omega_b$ is the angular velocity of the body. All vectors are taken in the laboratory frame of reference.
$P$ is positive if power is consumed.
We find that it is positive during most part of the takeoff (see Fig.~\ref{fig:drawing_voluntary}\textit{f}). Only at the reversals during the first two cycles, when the body velocity is still small, $P$ is slightly negative.
During the 2nd wingbeat, the mean body-mass specific aerodynamic power is equal to $P_b^*=P_{ave}/m=21$~W/kg.
Assuming that the muscles contribute to 30\% of the body mass, the mean muscle-mass specific aerodynamic power is equal to $P_m^*=P_{ave}/(0.3m)=69$~W/kg. 
The relative difference in the cycle averaged values between IGE and OGE is less than 0.5\%.


We conclude that the ground effect is unimportant for the voluntary takeoff, a result which is in agreement with \cite{Chen_Sun_2014}.
This is mainly a consequence of rapid acceleration during the first wingbeat cycle,
when the legs produce a large vertical force.
The main question of the next section is whether this scenario changes if the takeoff is slower and the insect remains near the ground for a longer time.
The rate of climb at the beginning of takeoff is controlled by the leg model stiffness coefficient $K_\ell^+$, and the horizontal velocity is controlled by the leg angle $\phi_\ell^+$.

\subsection{Slow takeoff}\label{sec:slow_vertical_takeoff}

This section describes a modified takeoff with the leg thrust coefficient decreased to $K_\ell^+=0.041$~N/m (see the second line in Table~\ref{tab:legs_params}).
Smaller $K_\ell^+$ results in less leg thrust and slower climb, compared to the natural voluntary takeoff.
Therefore we refer to this case as a `slow takeoff'.
Any further decrease of $K_\ell^+$ would result in a very close approach of the insect to the ground surface and, ultimately, to a collision
which would require special treatment. Modelling of such collisions could be an interesting topic for future research (see \cite{Wootton_etal_2003} for a review of structural modelling of insect wings, including impact modelling).
For a fruitfly, collisions between the wings and the ground may be undesirable because of the large wingbeat frequency and light wing structure.
Therefore, $K_l^+=0.041$~N/m is an interesting limiting case.

In the present numerical simulations, the computational domain size in $x$ and $y$ directions, the discretization grid step size and the penalization parameter are the same as in the previous section.
The domain size in the vertical direction $z$ is reduced to $6R$ because the insect gains much less altitude by the end of the simulation.

\begin{figure}
\begin{center}
\includegraphics{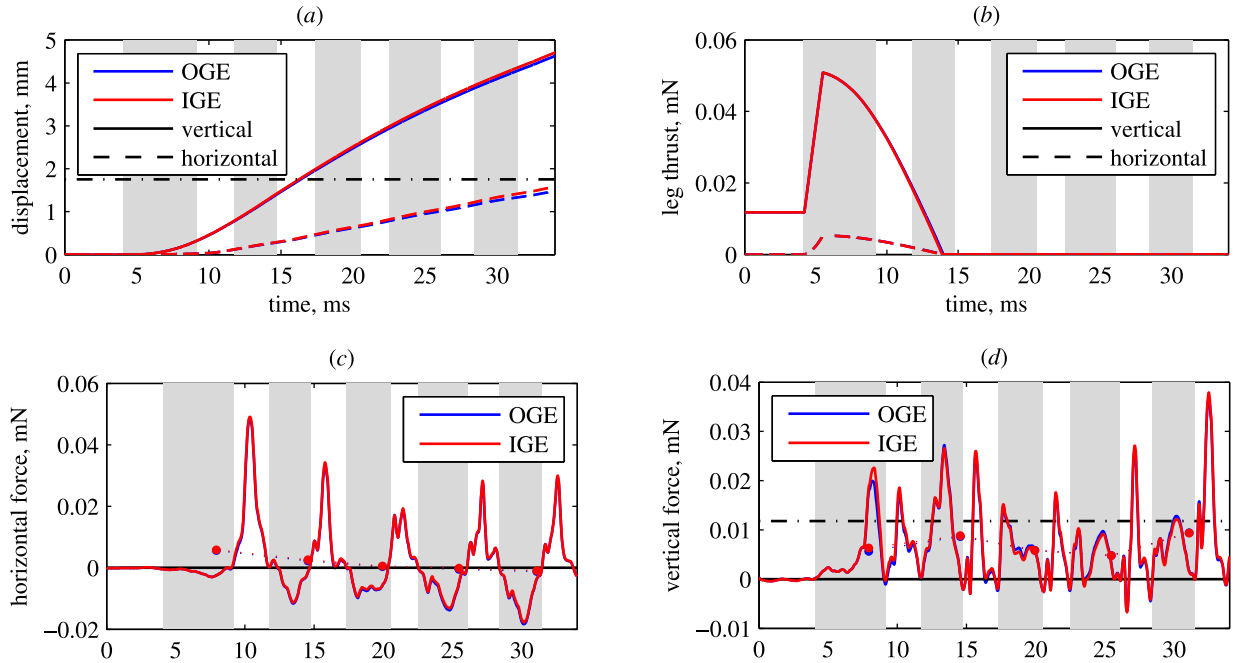}
\caption{\label{fig:traj_legs_forces_weakvert} {\bf Slow takeoff.} (\textit{a}) Vertical and horizontal displacement. Dash-dotted line indicates $z_c=R$. (\textit{b}) Components of the leg force.
(\textit{c}) horizontal and (\textit{d}) vertical components of the aerodynamic force. The black dash-dotted line in Fig. (\textit{d}) indicates the weight. Solid circles connected by dotted lines show wingbeat cycle averages. Note that, in Fig. (\textit{b}) and (\textit{c}), the lines for IGE and OGE almost coincide.}
\end{center}
\end{figure}

Fig.~\ref{fig:traj_legs_forces_weakvert}(\textit{a}) shows the displacement of
the body point of reference.
The rate of climb is about one third of its original value and the insect only gains 3.9~mm by the end of the 4th wingbeat cycle.
This is just slightly larger than the wing length $R$ (2.83 mm).
The displacement is slightly larger for IGE than for OGE in both directions, horizontal and vertical.
The time evolution of leg thrust is given in Fig.~\ref{fig:traj_legs_forces_weakvert}(\textit{b}).
There is no visible difference between the two cases. The peak of the vertical force is equal to $0.051~\rm{mN}$, which is about four times less than in the original voluntary takeoff discussed in section~\ref{sec:voluntary_takeoff}.

\begin{figure}
\begin{center}
\includegraphics{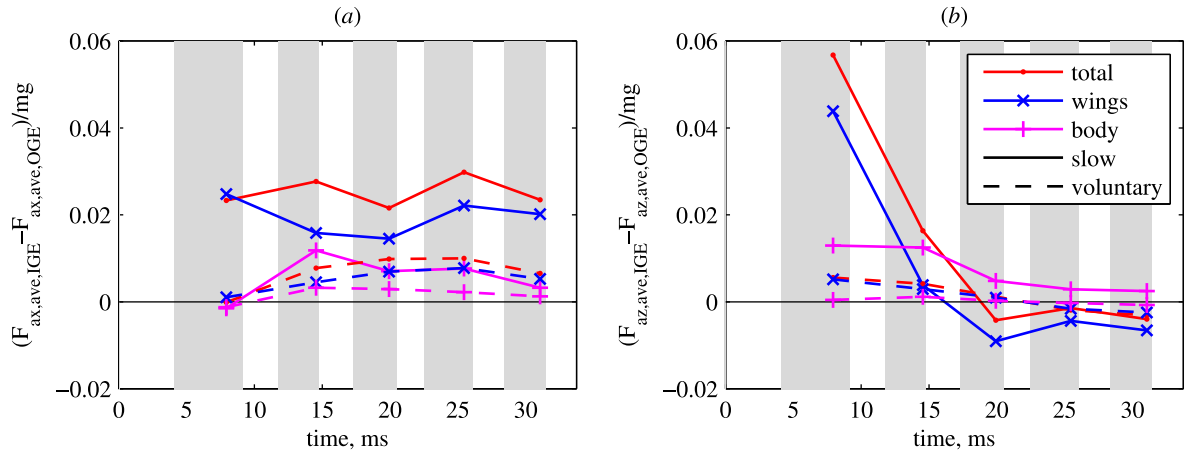}
\caption{\label{fig:forces_weakvert_ave} {\bf Slow and voluntary takeoffs.} The difference between the cases IGE and OGE, in terms of the wingbeat cycle averaged aerodynamic force normalized by the body weight (\textit{a}) horizontal and (\textit{b}) vertical components.
The forces acting on the wings and the body are shown separately. The total force, which is their sum, is also shown.}
\end{center}
\end{figure}

The time evolution of the instantaneous aerodynamic force, shown in Fig.~\ref{fig:traj_legs_forces_weakvert}(\textit{c,d}), is qualitatively similar to the voluntary takeoff case considered previously.
The difference between the cases OGE and IGE is negligible for the horizontal force (Fig.~\ref{fig:traj_legs_forces_weakvert}\textit{c}), but for the vertical force it reaches values as large as $0.0027~\rm{mN}$, i.e., 23\% of the weight (Fig.~\ref{fig:traj_legs_forces_weakvert}\textit{d}).
Fig.~\ref{fig:forces_weakvert_ave} shows the difference between the wingbeat averaged forces in the cases IGE and OGE, normalized by the weight.
The vertical force difference is shown in Fig.~\ref{fig:forces_weakvert_ave}(\textit{b}).
During the 1st wingbeat, the ground effect makes the total vertical force increase by almost 6\% of the weight (red line).
However, during the 2nd wingbeat, the extra force decreases to only 2\% of the weight.
During the 3rd, the 4th and the 5th wingbeats, the difference between the vertical forces IGE and OGE
is very small and negative. The increase of the vertical force during the first wingbeat is mainly due to the wings (blue line).
The extra force acting on the body is only about 1\% of the weight. However, later on,
the contribution of the body becomes important because it remains positive, whereas
for the wings it becomes negative. The horizontal force difference, shown in Fig.~\ref{fig:forces_weakvert_ave}(\textit{a}),
is positive, i.e., the propulsive force increases due to the ground effect by about 2\% of the weight, for all wingbeats.
The contribution of the body is up to 1\% of the weight.
For reference, Fig.~\ref{fig:forces_weakvert_ave} also shows the force differences during the voluntary takeoff.
They are all smaller than 1\%.


\begin{figure}
\begin{center}
\includegraphics{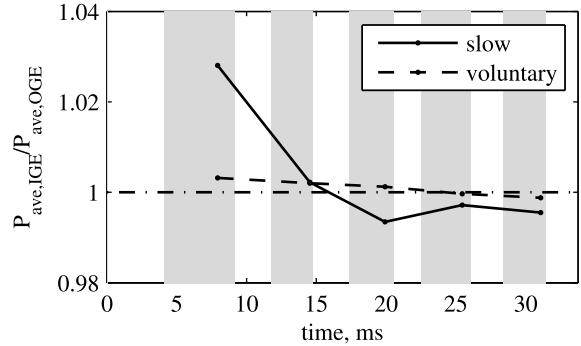}
\caption{\label{fig:power_cmp} {\bf Slow and voluntary takeoffs.} Aerodynamic power ratio IGE/OGE.}
\end{center}
\end{figure}

The aerodynamic power, in the cases IGE and OGE, is compared in Fig.~\ref{fig:power_cmp}.
The maximum difference is of about $3\%$ in magnitude for the slow takeoff, but less than $1\%$ for the voluntary takeoff.
Considering the slow takeoff, the insect consumes more power when operating in ground effect (IGE) during the first two wingbeat cycles, but less power during the subsequent cycles. Overall, we find that the differences in the power are small.

\subsection{Takeoffs with simplified kinematics}\label{sec:simple}

\begin{figure}
\begin{center}
\includegraphics{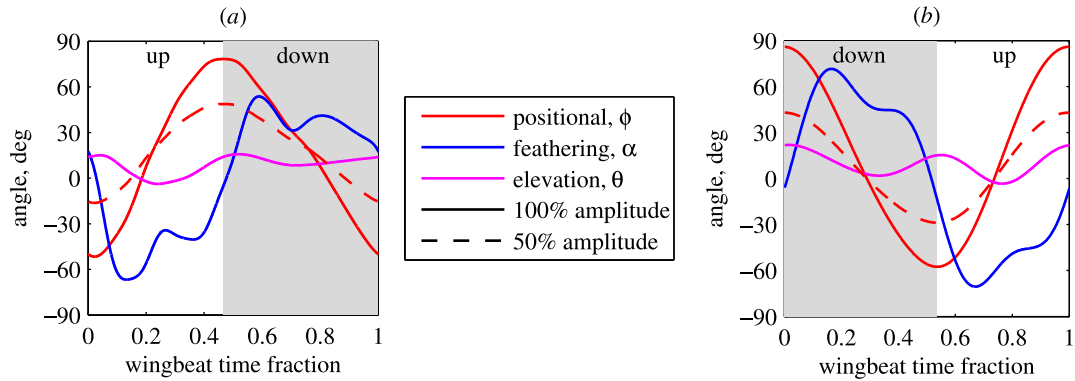}
\caption{\label{fig:kine} {\bf Wing kinematics.} (\textit{a}) Data obtained by periodization of the last wingbeat shown in \cite{Chen_Sun_2014}. It is used in section~\ref{sec:simple}. Also, in section~\ref{sec:hovering}, this kinematics is referred to as `P1' (`P' for `Periodic'). The cycle begins from the upstroke. (\textit{b}) Data adapted from \cite{Maeda_Liu_2013}. In section~\ref{sec:hovering}, it is referred to as `P2'. The cycle begins from the downstroke.}
\end{center}
\end{figure}

In the previous sections we noticed that the ground effect depends on the takeoff kinematics. We are mainly interested in the effects that might be generally applicable to a fruitfly sized insect takeoff. Therefore, in this section we consider parametric studies. They are performed using simplified periodic wing kinematics. The time evolution of the three wing angles over one wingbeat period, obtained by periodization of the last wingbeat in \cite{Chen_Sun_2014}, is shown in Fig.~\ref{fig:kine}(\textit{a}).
The wingbeat frequency is equal to $f=210$~Hz.  
This takeoff mode can be relevant to MAVs, for which the wing kinematics and the body angle during takeoff do not change as much as for the fruitfly (see, e.g, \cite{Chirarattananon_etal_2013}).
The leg strength parameter $K_{\ell}^+$ is varied, resulting in a variation of the takeoff rate of climb $V_{t.o.}$.
The body angle is constant and equal to $\beta=46.3^\circ$, the anatomical stroke plane angle is equal to $\eta=32^\circ$. 
In these computations, we use $L_x=L_y=4R$, $L_z=6R$, $N_x=N_y=512$ and $N_z=768$, corresponding to more than 200 million grid points. The penalization parameter is equal to $\varepsilon = 2.5\cdot10^{-4}$.

\begin{figure}
\begin{center}
\includegraphics{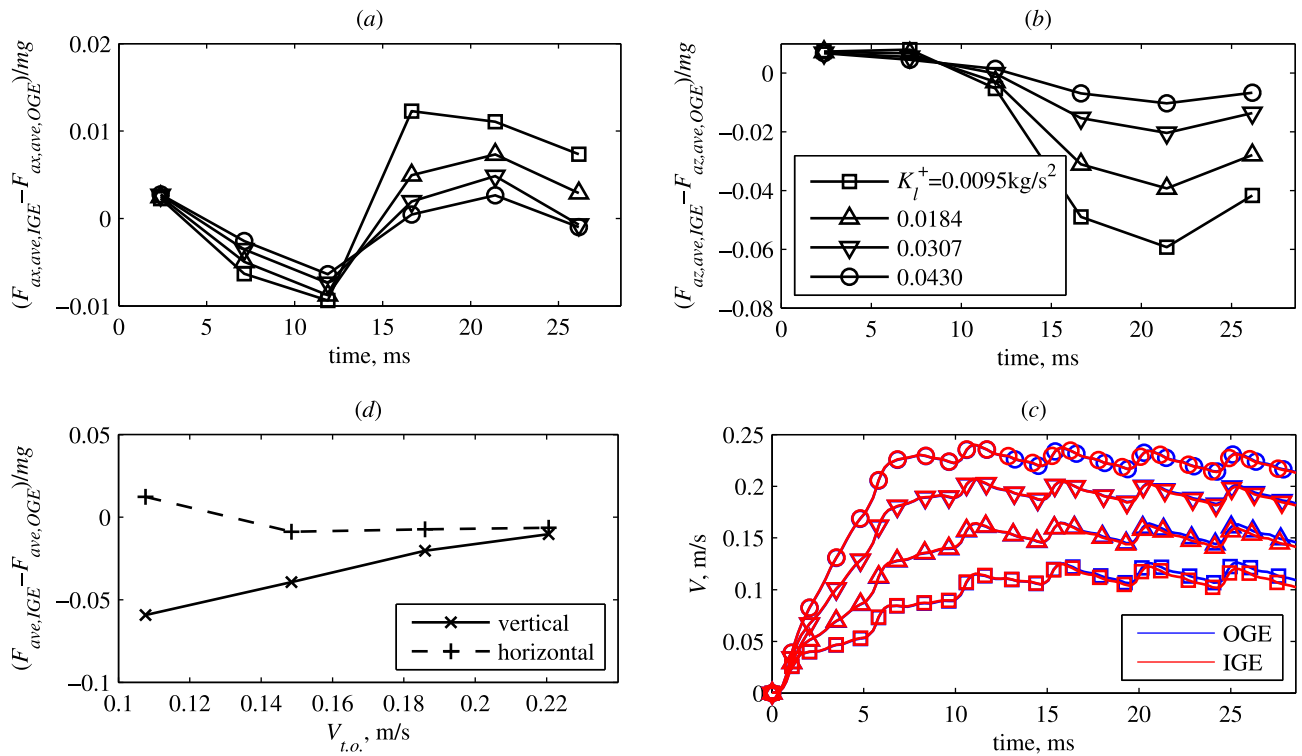}
\caption{\label{fig:simplified} {\bf Takeoffs with simplified kinematics.} (\textit{a}) Horizontal and (\textit{b}) vertical difference between the wingbeat-averaged force IGE and OGE, normalized to the insect weight.
(\textit{c}) Vertical velocity of the body point of reference (rate of climb) versus time.
(\textit{d}) Maximum normalized force difference versus takeoff rate of climb at the moment when the legs lose contact with the ground.}
\end{center}
\end{figure}

Smaller $K_l^+$ implies smaller rate of climb (Fig.~\ref{fig:simplified}\textit{c}) which leads to a more significant ground effect (Fig.~\ref{fig:simplified}\textit{d}).
A striking feature of Fig.~\ref{fig:simplified}(\textit{b}) is
a significant decrease of the vertical force during the 4th, 5th and 6th wingbeats, by up to 6\%.
The horizontal force varies slightly, by about 1\% (see Fig.~\ref{fig:simplified}\textit{a}).

\begin{figure}
\begin{center}
\includegraphics{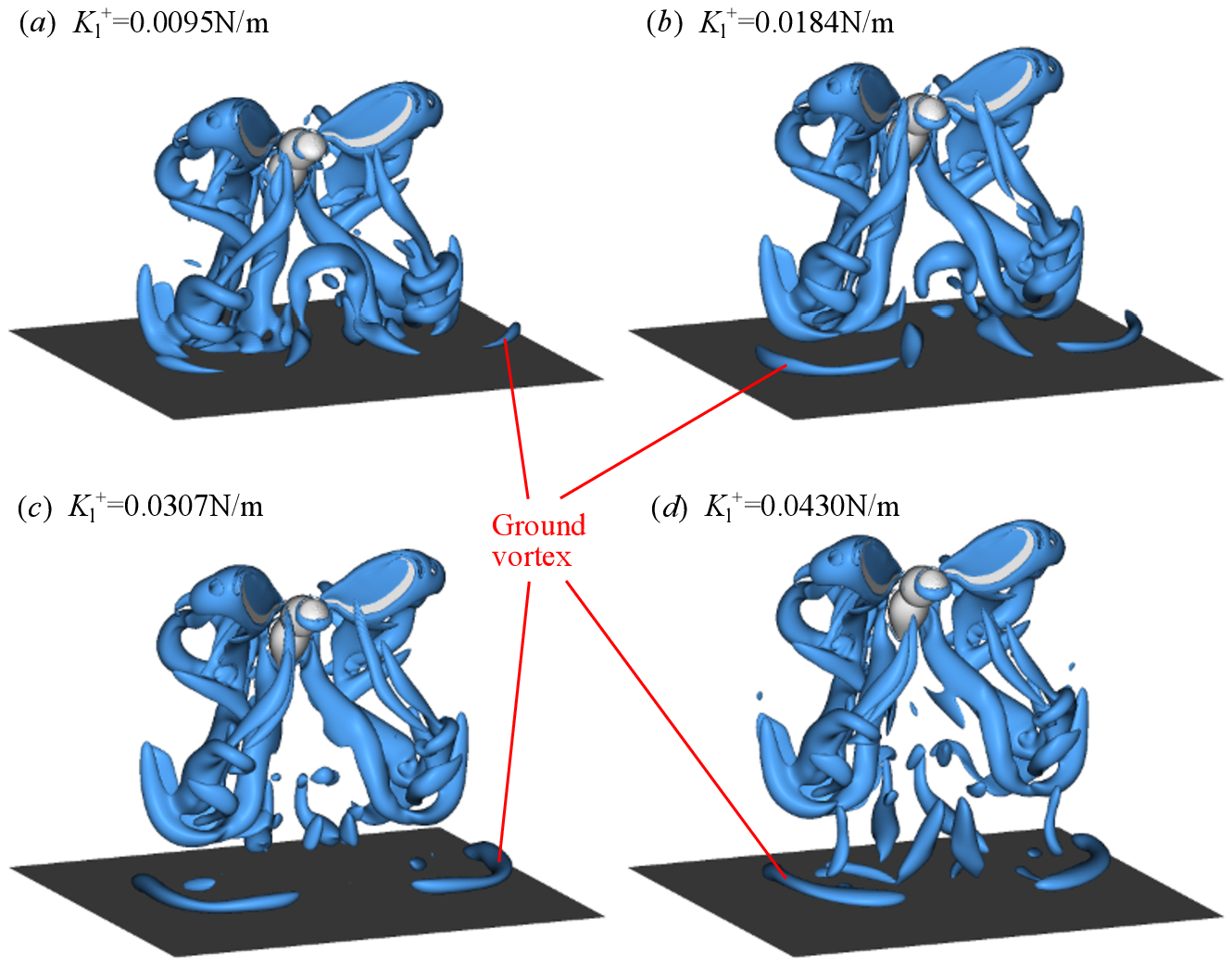}
\caption{\label{fig:simplified_visu} {\bf Flow visualization at the end of the 5th wingbeat ($t=23.78$~ms) for the simplified kinematics cases.} Iso-surfaces $Q/f^2=15$ are shown for 4 different takeoffs IGE: (\textit{a}) $K_l^+=0.0095$~N/m;
(\textit{b}) $K_l^+=0.0184$~N/m;
(\textit{c}) $K_l^+=0.0307$~N/m;
(\textit{d}) $K_l^+=0.0430$~N/m.}
\end{center}
\end{figure}


The decrease in vertical force on the wings found in the IGE cases compared to OGE (Fig.~\ref{fig:simplified}\textit{b}) is an adverse ground effect which
may be the result of complex wing-wake-ground interactions that depend on wing and body kinematics.
Adverse ground effects have been reported previously for fixed-wing aircraft \cite{Jones_PhD_2005}, but they consist in increased drag together with increased lift.
However, for a two-dimensional ellipse with normal hovering kinematics \cite{Gao_Lu_2008}, the mean vertical force decreases when the height from the ground is between approximately $1.5D$ and $4D$, where $D$ is the chord length of the ellipse.
For flapping wings, an adverse ground effect was found by Quinn et al. \cite{Quinn_etal_2014}. They considered an airfoil undergoing pitch
oscillations in a closed-loop water channel with prescribed free-stream velocity. Such a configuration represents a section of a bird wing in forward flight or a fish fin.
Even if the pitching motion was symmetric, the proximity of the ground broke the symmetry of the flow. Thus, the airfoil produced non-zero lift.
The lift was positive if the distance to the ground was less than 40\% of the chord length, but it became negative at larger distances, such that the lift force pulled the
airfoil towards the ground.
Nevertheless, the extra propulsive force due to the ground effect was positive in all cases.
Note, however, that the flows considered in \cite{Gao_Lu_2008} and \cite{Quinn_etal_2014} are effectively two-dimensional.

In the present work, importantly, we find an adverse ground effect in a three-dimensional configuration, which has not been previously recognized.
The wingbeat cycle averaged vertical force of the wings in the ground effect is slightly larger during the first two cycles, but,
as the insect flies away from the ground, the vertical force in the case IGE becomes less than that in the case OGE.

Fig.~\ref{fig:simplified_visu} shows a comparison
of the wake at the end of the 5th wingbeat cycle ($t=23.78$~ms) in the 4 different takeoffs IGE with different values of $K_l^+$.
It corresponds to the maximum decrease of the vertical force.
There are noticeable differences between the vortices when the takeoff is fast and when it is slow.
The part of the wake that approaches the ground deforms when it impinges on the ground.
It then rolls up in a pair of vortex rings. Similar ``ground vortices''
are known in the context of helicopter rotor aerodynamics.
In each of the 4 cases shown in Fig.~\ref{fig:simplified_visu},
they have different strength and position with respect to the wings.
Therefore, they induce the downwash of different strength.

\begin{figure}
\begin{center}
\includegraphics{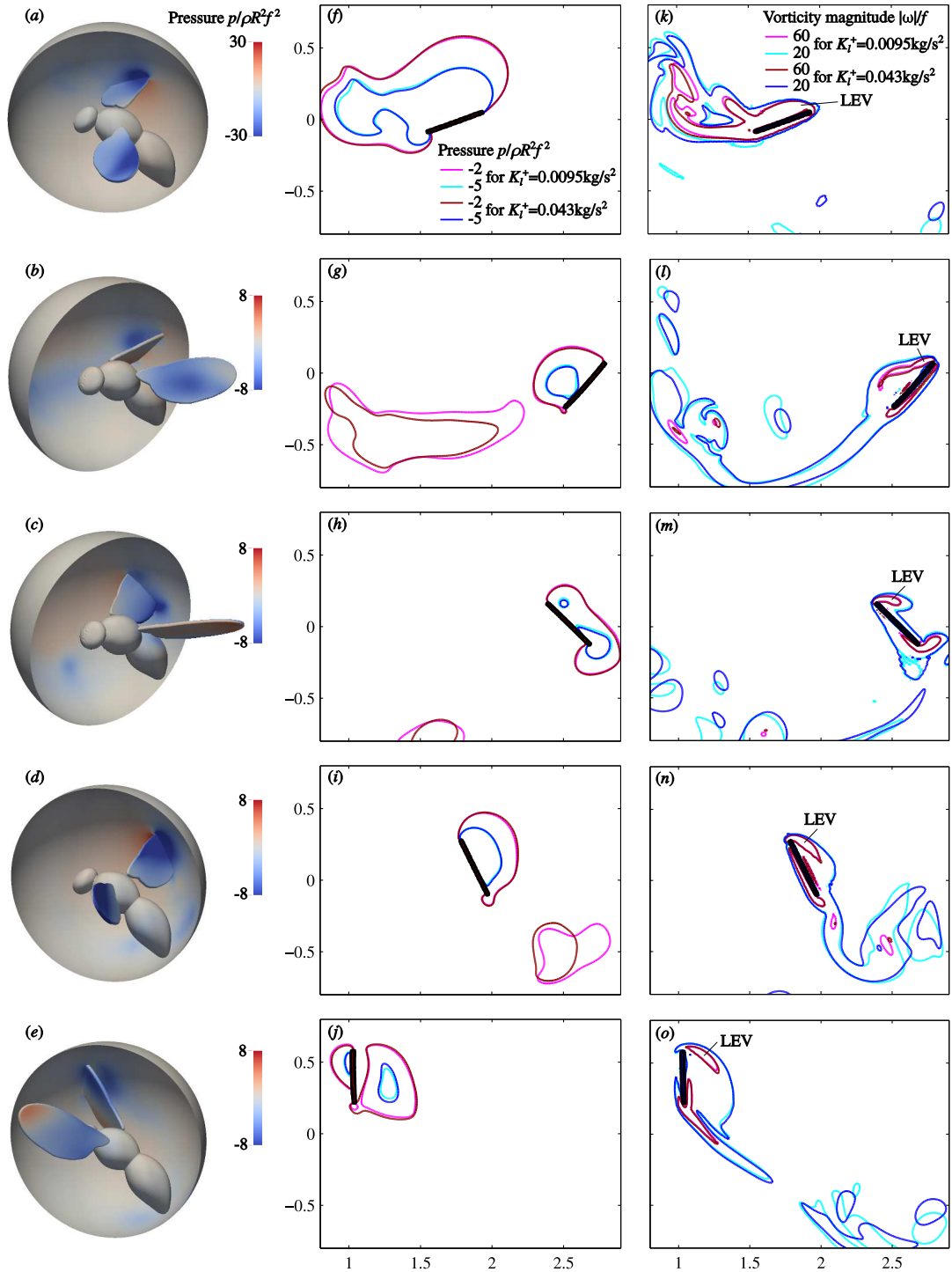}
\caption{\label{fig:lev_visu} {\bf Visualization of the leading-edge vortex during the 6th wingbeat for the simplified kinematics cases, IGE.}
Left column (\textit{a}-\textit{e}) shows the dimensionless pressure distribution over the surface of the model and over a semisphere of radius $0.9R$
around the body point of reference. Middle column (\textit{f}-\textit{j}) shows the pressure iso-contours for two different takeoffs: $K_l^+=0.0095$~N/m and $0.0430$~N/m.
Right column (\textit{k}-\textit{o}) compares iso-contours of the dimensionless vorticity magnitude for the same two takeoffs.
Time instants are $t=24.73$, $25.68$, $26.63$, $27.58$ and $28.53$~ms ($tf=5.2$, $5.4$ $5.6$, $5.8$ and $6$, where $f$ is the wingbeat frequency).
}
\end{center}
\end{figure}

A detailed view of the flow near the wings is presented in figure~\ref{fig:lev_visu}.
It shows the pressure and the vorticity magnitude during the 6th wingbeat for two IGE cases with different $K_l^+$.
Five time instants are visualized in five rows, respectively. The left column shows the pressure distribution
over the surface of the insect, as well as over a semisphere of radius $0.9R$ centred the body reference point, for $K_l^+=0.0095$~N/m.
Figure~\ref{fig:lev_visu}(\textit{a}) is at $t=24.73$~ms, during upstroke. The dark blue area
near the leading edge that expands towards the wing tip is the trace of the leading-edge vortex (LEV), similar to the one discussed in \cite{Aono_etal_2008}.
The LEV at the downstroke is evident in figure~\ref{fig:lev_visu}(\textit{d}) at $t=27.58$~ms. The pressure distributions during the reversals
are more complex (figures~\ref{fig:lev_visu}\textit{b}, \textit{c} and \textit{e}).

The middle column of figure~\ref{fig:lev_visu} shows the pressure iso-contours sampled on the surface of the semisphere.
Two cases are compared, $K_l^+=0.0095$~N/m and $0.0430$~N/m.
This choice of $K_l^+$ corresponds to the largest difference in the vertical force (see figure~\ref{fig:simplified}\textit{b}).
For each of the cases, two isolines are drawn, $p/\rho R^2 f^2=-2$ and $-5$.
There is a large difference between the contours for different $K_l^+$ in the far wake (wake far from the wings).
However, in the near wake of a wing including the LEV, the difference is much smaller,
which is consistent with the force changing by only a few per cent.

The iso-contours of the vorticity magnitude are compared in the right column.
Here again, despite significant differences in the far wake,
the LEV contours virtually overlap for $K_l^+=0.0095$~N/m and $0.0430$~N/m.
This shows that the ground effect has almost no influence on the dynamics of the LEV.

\subsection{Ground effect in hovering flight}\label{sec:hovering}

In this section, we simplify the kinematics even further. We consider hovering with the insect body being fixed.
The flight dynamics solver is not used in this case.
The distance from the body centre to the ground is equal to $0.48R$ (where $R$ is the wing length) for hovering in ground effect (IGE)
and $2.4R$ for hovering out of ground effect (OGE).
The body pitch angle and the anatomical stroke plane angle are both constant and equal to $55^\circ$, such that the stroke plane is horizontal. 
The wing kinematics is the same as in the previous section, see Fig.~\ref{fig:kine}(\textit{a}). We denote it as `P1' kinematics. The wingbeat frequency is equal to $f=218$~Hz.
The first wingbeat starts from the upstroke, as done in \cite{Chen_Sun_2014}.

In these numerical simulations we are interested in the long-time evolution of the aerodynamic forces,
which after the initial transient eventually reach a periodic state.
Most of the results known from the helicopter rotor theory \cite{Leishman_2006} are obtained in reference to the periodic state,
while the takeoffs considered in the previous sections of this paper (the slow takeoffs, in particular) last only for a few wingbeats.
Therefore, it is instructive to consider the time evolution of the aerodynamic forces during hovering from $t=0$ until the time when the periodic state is
reached.

Since the time span of the numerical simulations presented in this section is large, it is necessary to increase the domain size in the horizontal directions.
We set $L_x \times L_y \times L_z = 8R \times 8R \times 4R$, where $z$ is the vertical direction.
The number of grid points is $N_x \times N_y \times N_z = 864 \times 864 \times 432$.
The penalization parameter is equal to $\varepsilon = 2.5\cdot10^{-4}$.

\begin{figure}
\begin{center}
\includegraphics{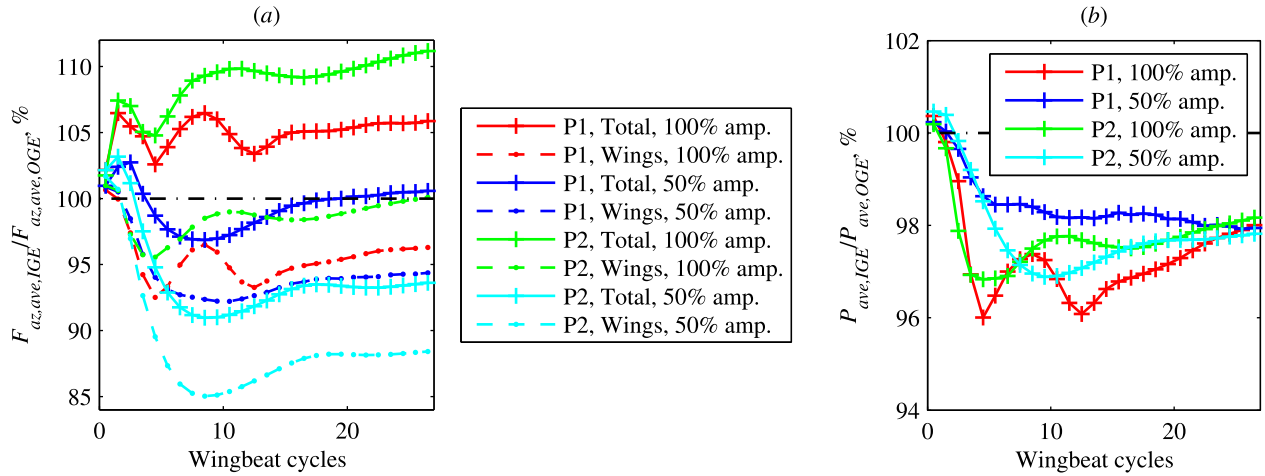}
\caption{\label{fig:frel_hovering} {\bf Hovering flight.} (\textit{a}) The ratio of wingbeat averaged vertical force IGE to OGE, $F_{z,ave,IGE}/F_{z,ave,OGE}$.
(\textit{b}) The ratio of wingbeat averaged power IGE to OGE, $P_{ave,IGE}/P_{ave,OGE}$.}
\end{center}
\end{figure}

The quantity of interest is the ratio of the wingbeat averaged forces, IGE to OGE: $F_{z,ave,IGE}/F_{z,ave,OGE}$.
This quantity is shown in Fig.~\ref{fig:frel_hovering}(\textit{a}).
The red solid line with ``$+$'' symbols corresponds to the total vertical force ratio.
As already noticed in \cite{Maeda_Liu_2013}, the vertical force during hovering in ground effect, $F_{z,IGE}$,
reaches its periodic state significantly later than during hovering out of ground effect, $F_{z,OGE}$.
Therefore, the ratio of their wingbeat averages, $F_{z,ave,IGE}/F_{z,ave,OGE}$ converges slowly with the number of wingbeats.
It oscillates between $102\%$ and $107\%$. After 27 wingbeats it reaches $106\%$. 

A similar comparison for the force generated by the wings is shown with a red dot-dashed line in Fig.~\ref{fig:frel_hovering}(\textit{a}).
It was calculated by integration of the distributed forces over the wings only, in the same numerical simulations. Therefore, the aerodynamic interaction between the body and the wings is included.
This force ratio drops from $100.6\%$ to $92.5\%$ during the first 6 wingbeats, oscillates and then increases to $96.3\%$.

\begin{figure}
\begin{center}
\includegraphics{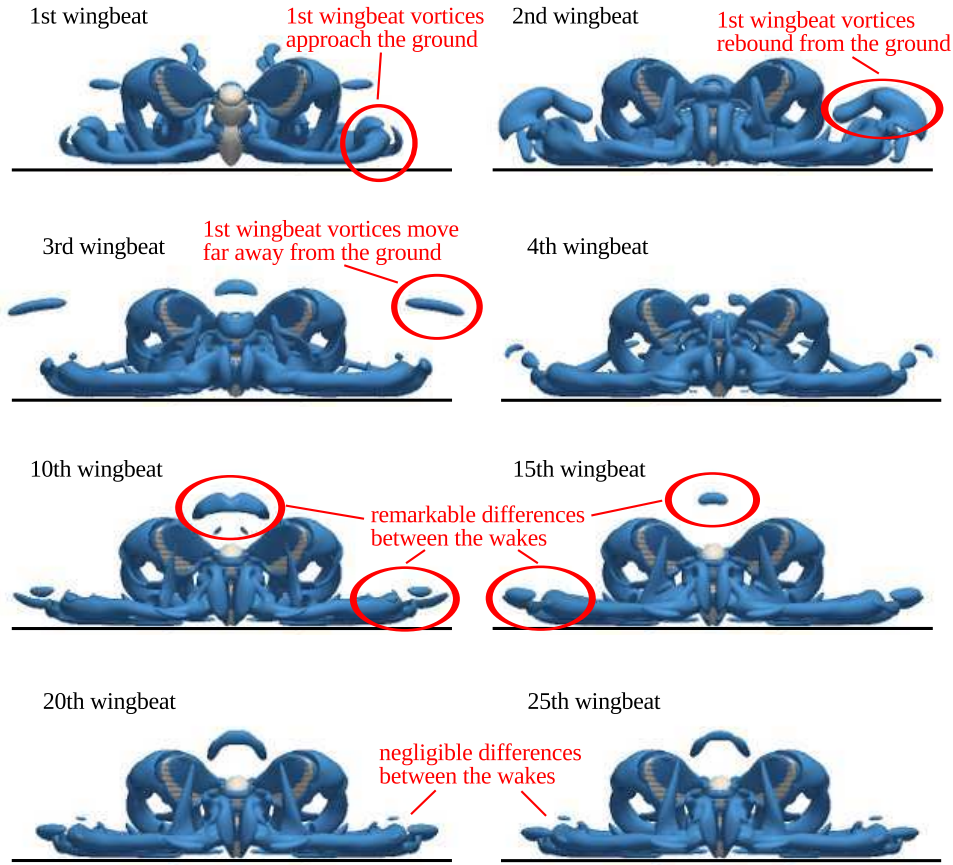}
\caption{\label{fig:wake_hovering} {\bf Time evolution of the wake during hovering.} Iso-surfaces of the $Q$-criterion, $Q/f^2=15$, are shown at the end of the downstroke. `P1' kinematics with 100\% wingbeat amplitude.} 
\end{center}
\end{figure}

The time evolution of the wake vortices generated by the insect is shown in Fig.~\ref{fig:wake_hovering}.
The visualized time instants correspond to the end of the 1st, 2nd, ..., 25th downstroke.
There are significant differences between the positions of the vortices over the first four time instants.
The first wingbeat generates very strong vortex rings, that collide with the ground. Then they rebound during the second wingbeat,
and parts of them moving upwards are still visible during the third wingbeat.
The downwash produced by these vortices influences the nearer wake dynamics and it is likely to be responsible for the decrease of the vertical force
during the first few wingbeats.
After the 10th wingbeat, the wake approaches its quasi-periodic state.
There are almost no visible differences between the visualizations at the end of the 20th wingbeat and at the end of the 25th wingbeat.

The pair of numerical simulations (cases IGE and OGE) that we have discussed in the above paragraphs leads to the following conclusions.
\begin{enumerate}
\item Over the first 27 wingbeats, $F_{z,ave,IGE}/F_{z,ave,OGE}$ varies within about $5\%$ for the total force and $9\%$ for the wings force.
\item The wing generates less vertical force in the case IGE than in the case OGE (adverse ground effect).
\item The body makes an important contribution to the total vertical force when operating IGE, which results in the excess total vertical force (positive ground effect).
\end{enumerate}
These conclusions are, of course, only valid for the particular wing shape and kinematics used in the simulation.
Periodic flapping is only an approximation to the real insect wing motion, which varies from one wingbeat to another,
and depends on many different conditions.
To determine the effect of all existing fruitfly wing kinematics is beyond the reach of our numerical simulations.
However, it is useful to compare a few different cases.

We carried out numerical simulations with the wing kinematics used in \cite{Maeda_Liu_2013} (abbreviated as `P2' in the figures).
Note that, in this case, the first wingbeat begins from the downstroke, as shown in Fig.~\ref{fig:kine}(\textit{b}).
The wingbeat frequency is the same, $f=218$~Hz.
The results of these numerical simulations are shown in Fig.~\ref{fig:frel_hovering} with green lines.
They are qualitatively similar to the previously shown `P1' case, but the values are systematically larger.
$F_{z,ave,IGE}/F_{z,ave,OGE}$ reaches 111\% for the total force and 100\% for the wings,
such that there is no adverse ground effect after the periodic state is established.

The adverse ground effect is rarely encountered in the aircraft or rotorcraft aerodynamics literature.
However, in the context of flapping wings, it is not unusual.
In the two-dimensional numerical simulations \cite{Gao_Lu_2008}, a U-shape profile of the force ratio $F_{z,ave,IGE}/F_{z,ave,OGE}$ versus $h/c$ was found, where $h$ is the distance from the wing centre to the ground and $c$ is the wing chord. $F_{z,ave,IGE}/F_{z,ave,OGE}$ was greater than $100\%$ for $h/c<1.5$,
but less than $100\%$ for $h/c>1.5$, and the minimum ratio was of about $54\%$.

In the three-dimensional model considered in the present paper, it is possible to partially reduce the three-dimensional effects by decreasing the wing beat amplitude.
One can then expect the adverse ground effect to be amplified. Indeed, this is what we find by decreasing the wingbeat amplitude by a factor of 2
(by rescaling the positional angle shown in Fig.~\ref{fig:kine} such that the amplitude of the positional angle is halved but the mean positional angle is unchanged, and the wingbeat frequency remains unchanged).
The results are shown in Fig.~\ref{fig:frel_hovering} with blue lines.
Now we have $F_{z,ave,IGE}/F_{z,ave,OGE}<100\%$ at intermediate times, for wings force and for the total force. The final periodic state produces a very slight excess of the total vertical force (less than $1\%$). Note that, in this reduced-amplitude case, the total force ratio decreases more than the wing force ratio. This indicates, not surprisingly, that the fountain effect becomes weaker when the wingbeat amplitude is reduced.
Similar computations with the `P2' kinematics show the same trend with an even larger decrease of $F_{z,ave,IGE}/F_{z,ave,OGE}$.

The wingbeat averaged aerodynamic power ratio $P_{ave,IGE}/P_{ave,OGE}$ is shown in Fig.~\ref{fig:frel_hovering}(\textit{b}).
Its variation is smaller than the variation of the force, and the computations suggest that its long-time limit is between $97\%$ and $99\%$, in all cases that we have considered.
The shape of the time evolution profiles of the power ratio is approximately similar to the time profiles of the wings vertical force ratio.
This means that a local decrease of the vertical force ratio is accompanied by a decrease of the power ratio.
Therefore, if the kinematics of the wings operating in ground effect is adjusted such that $P_{ave,IGE}/P_{ave,OGE}=100\%$ at any time, the force ratio $F_{z,ave,IGE}/F_{z,ave,OGE}$
is likely to increase. Among other factors, the feathering angle is very likely to change passively, when in ground effect, due to compliance of the wing \cite{Ishihara_etal_2009,Tanaka_etal_2011,Ishihara_etal_2014}. Such effects would need further investigation.

\section{Conclusions}\label{sec:conclusions}

The aerodynamic ground effect in fruitfly sized insect takeoff has been studied numerically using high performance computing.
The three-dimensional incompressible Navier--Stokes equations
were solved using a pseudo-spectral method with volume penalization using the \textsc{FluSi} open source code \cite{Engels_etal_2015,Kolomenskiy_Schneider_2009,Kolomenskiy_etal_11b},
in order to obtain the flow field and the aerodynamic forces acting on the insect.
The takeoff trajectories were calculated using a simple
flight dynamics solver that accounts for the body weight, inertia, and the legs thrust.
A series of computations has been carried out to explore the parametric space of the model.
A natural voluntary takeoff of a fruitfly, modified takeoffs
with different kinematics and leg model parameters, and hovering flights have been compared.

We found that the ground effect during the natural voluntary takeoff is negligible.
The wingbeat averaged forces only differ by less than 1\% of the weight.
The aerodynamic power differs by less than 0.5\%.

In the modified takeoffs, we decreased the leg strength.
As a consequence, the rate of climb decreased and the ground effect became significant.
Surprisingly, the vertical force did not always increase.
It even dropped in some of the cases that we considered.
This is an unsteady effect related to the vortex rings bouncing off the ground surface.

To better understand the mechanism of the adverse ground effect,
we considered hovering near a flat ground surface,
being the limiting case of zero rate of climb.
In that case, the fountain effect produced a large upward force on the insect's body.
The net ground effect was therefore positive. However, the aerodynamic force
acting on the wings in ground effect was sometimes less than when the wings operate out of ground effect.
The most significant decrease was observed during the first 15 wingbeats. Note that this is a much longer time period than a typical takeoff.
At long time hovering, the effect was either positive or negative, depending on the wings kinematics.

The parameter space in the takeoff problem is very large.
In the present study, we focused on the legs thrust and wing kinematics.
However, the aerodynamic ground effect may also be sensitive to the Reynolds number,
because the structure of the wake at high $Re$ is significantly different from that at low $Re$.
Since high Reynolds number computations are costly,
they are beyond the scope of the present study,
but it is an important question for future research.







\appendix

\section{Numerical validation of the ground plane modelling using the volume penalization method.}
\label{sec:validation}

A numerical study of the ground effect during hovering flight
was carried out by Gao and Lu \cite{Gao_Lu_2008} in the two-dimensional approximation.
In this section, we compare with some of their results.

\begin{figure}
\begin{center}
\includegraphics[scale=0.6]{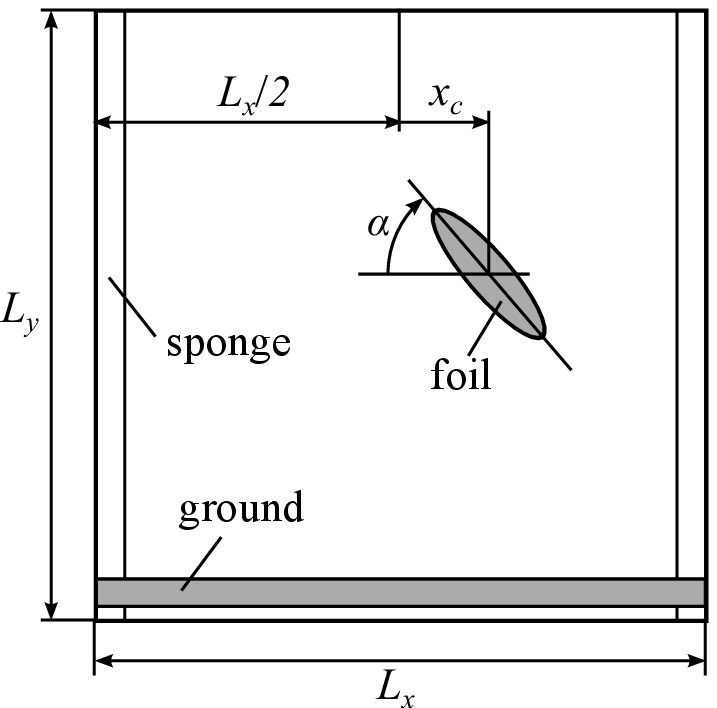}
\caption{\label{fig:domain_2d}{\bf Schematic drawing of the setup.}}
\end{center}
\end{figure}

All quantities in \cite{Gao_Lu_2008} are presented in a non-dimensional form, and we follow the same conventions.
The wing cross-section is an ellipse, as schematically shown in Fig.~\ref{fig:domain_2d}.
Its major axis (chord length) is $c=1$, and its minor axis is equal to 0.25.
The motion of the ellipse centre in the horizontal direction ($x$ direction) is given by
\begin{equation}
x_c(t) = A_m \cos(2 \pi t / T),
\label{eq:ellipse_kine_x}
\end{equation}
where $A_m = 1.25$, and we fix the dimensionless stroke time period to $T=2\pi A_m$.
The vertical coordinate of the ellipse centre is constant in time.
The angle between the major axis and the horizontal axis varies according to
\begin{equation}
\alpha(t) = \alpha_0 - \alpha_m \sin(2 \pi t / T),
\label{eq:ellipse_kine_alpha}
\end{equation}
where $\alpha_0 = 90^\circ$ and $\alpha_m = 45^\circ$.

The ground surface
is horizontal ($x$ direction), so that the distance $D$ between the ellipse centre and the ground
remains constant in time. In the present study, we vary this parameter between 1 and 6.

The dimensionless density of the fluid is $\rho=1$.
The dimensionless kinematic viscosity is equal to $\nu=10^{-2}$ yielding the Reynolds number
\begin{equation}
Re = \frac{U c}{\nu} = 100, \quad\quad \mathrm{where} \quad U = \frac{2\pi A_m}{T} = 1.
\label{eq:reynolds_2d}
\end{equation}

Our method solves the three-dimensional incompressible Navier--Stokes equations.
The two-dimensional flow is therefore modelled by imposing the initial and boundary conditions
constant in the direction perpendicular to the flow plane ($z$ direction).
The domain size in this direction is $L_z=1$. 

In the $xy$ plane, the computational domain size is equal to $L_x \times L_y = 12 \times 12$.
The number of grid points is equal to $N_x \times N_y = 512 \times 512$ (\emph{low resolution}) or $1024 \times 1024$ (\emph{high resolution}). In the low resolution simulations, the grid step size $\Delta x \approx 0.0234$ is comparable
to the lowest value reported in \cite{Gao_Lu_2008} ($\Delta x=0.025$).


The ground is modelled as a solid layer of width 0.2. Its top surface is at distance $D$ below the centre of the ellipse.
Smoothing of the penalization mask function (erf, $3\Delta x$ inwards and $3\Delta x$ outwards, see  \cite{Engels_etal_2014}) is only applied to the ellipse, not to the ground.

The penalization parameter, both for the ellipse and for the ground, is equal to $\varepsilon = 10^{-3}$ in the low-resolution simulations and $2.5 \cdot 10^{-4}$ in the high-resolution simulations.
A `vorticity sponge' forcing term \cite{Engels_etal_2014} is introduced in the momentum equation equation
in order to weaken the effect of the periodic boundary conditions in $x$. It is applied in two vertical layers of thickness equal to $32 \Delta x$, and its penalization parameter is equal to $\varepsilon_{sponge} = 0.1$.

\begin{figure}
\centering
\includegraphics[scale=0.65]{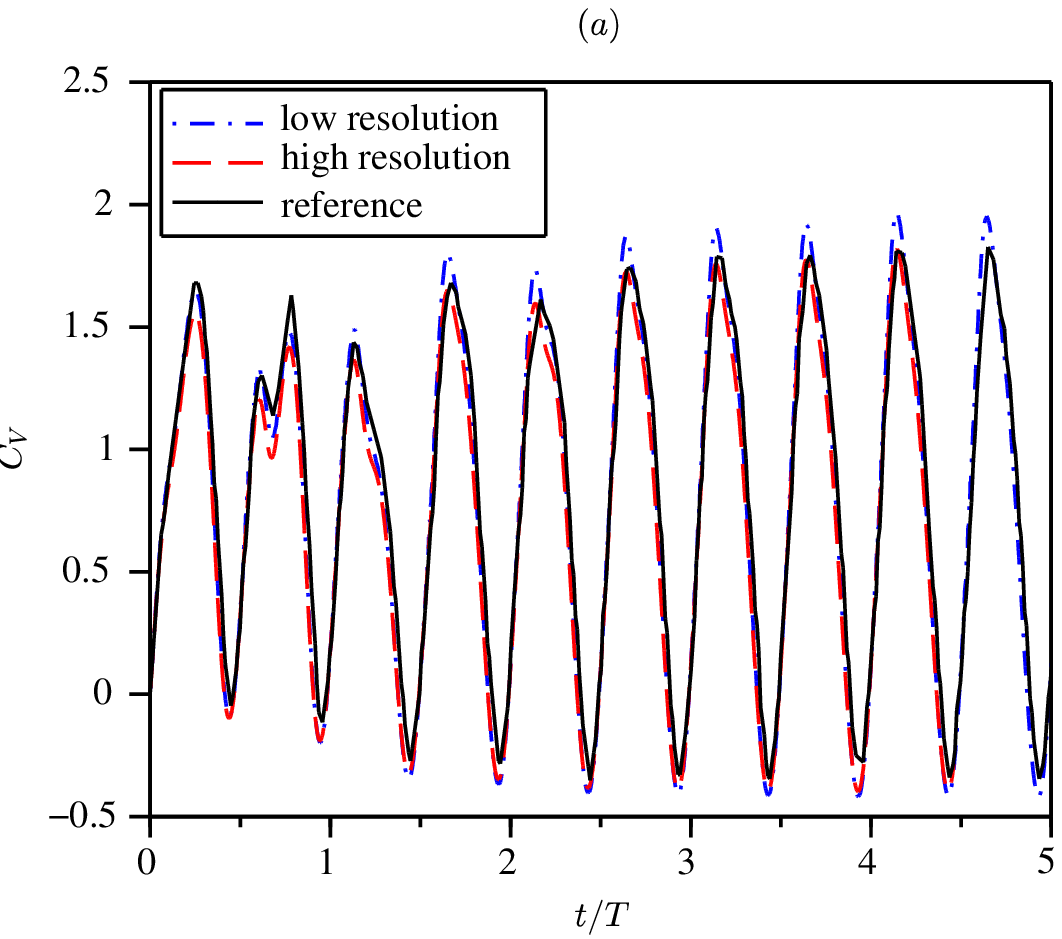}
\includegraphics[scale=0.65]{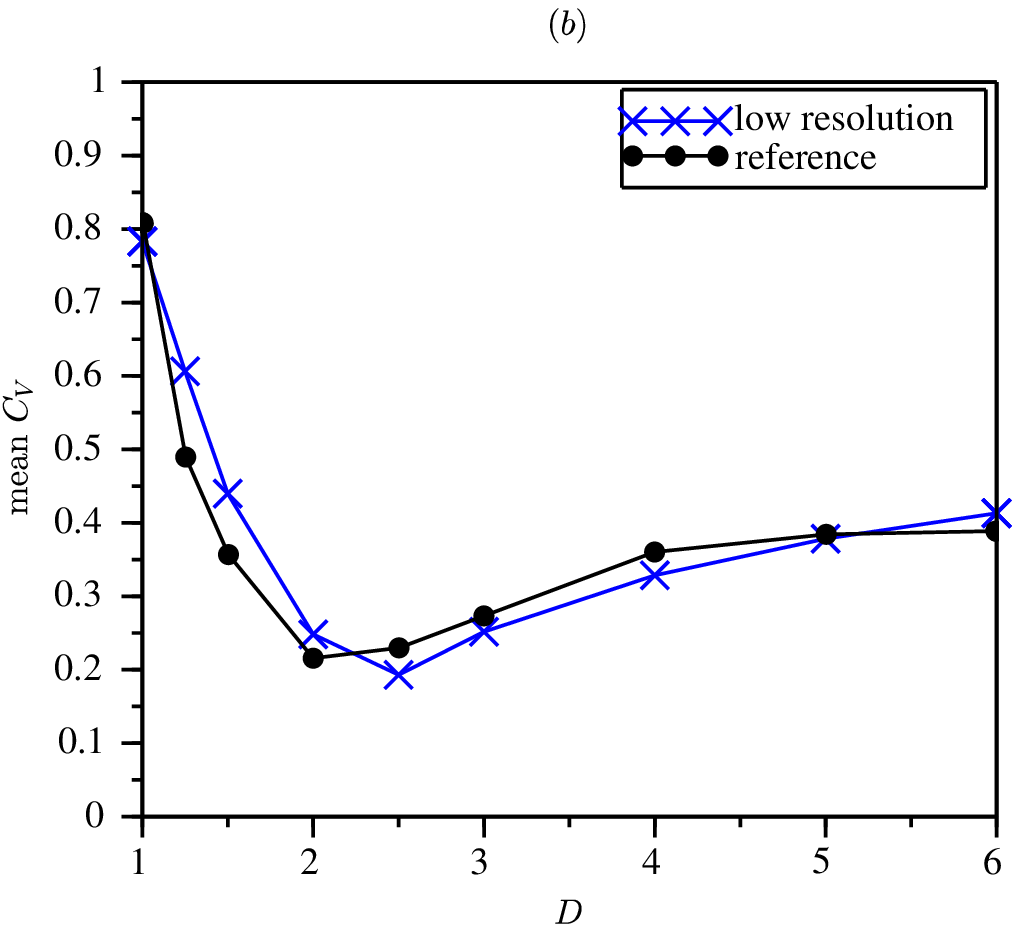}
\caption{{\bf Results of the numerical simulations.} (\textit{a}) Time evolution of the vertical force coefficient $C_V$ at the distance from the ground $D=1$.
(\textit{b}) Mean vertical force coefficient versus distance from the ground.}\label{fig:2d_results}
\end{figure}

The nearest distance to the ground in \cite{Gao_Lu_2008} is $D=1$. Fig.~\ref{fig:2d_results}(\textit{a}) displays the time evolution of the vertical force coefficient $C_V$, obtained by normalizing the vertical force $F_V$,
\begin{equation}
C_V = \frac{F_V}{0.5 \rho U^2 c L_z}.
\label{eq:force_coefs_2d}
\end{equation}
where $U = 2\pi A_m/T$. In the figures, the time $t$ is normalized to the stroke period $T$.
After $t/T=1$, oscillations of $C_V$ are mainly described by the second harmonic.
They become apparently periodic after $t/T=4$.

A series of simulations has been carried out in order to determine how the time averaged force coefficients depend on $D$.
Their parameters correspond to the low resolution, as defined above.
As shown in Fig.~\ref{fig:2d_results}(\textit{b}), the minimum of $C_V$ is observed in our simulations at about the same $D$ as in \cite{Gao_Lu_2008}.
We conclude that, in this two-dimensional validation case, our results are in reasonable agreement with the reference \cite{Gao_Lu_2008}.

\section{Influence of the domain size in the vertical direction.} \label{sec:domain_size}

In our numerical simulations of takeoffs, the insect approaches the top of the computational domain
as time increases. The simulations have to be stopped before the finite domain size begins to influence the results significantly.
To quantify this effect, we compare the aerodynamic forces and power in the voluntary takeoff, computed using
two different vertical domain sizes, $L_z=6R$ and $L_z=8R$.
We consider the OGE case in which the bee is nearer to the top of the domain.

\begin{figure}
\begin{center}
\includegraphics[scale=1.0]{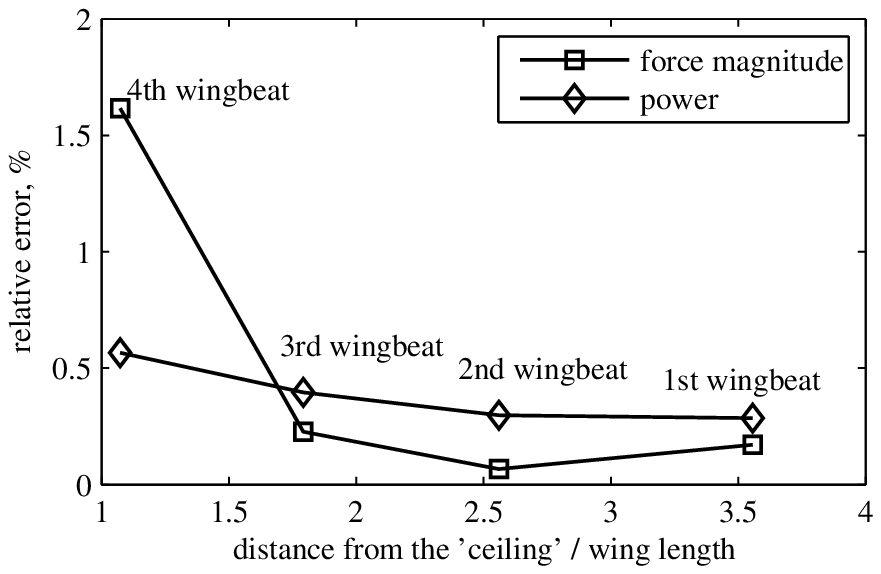}
\caption{\label{fig:domain_size_cmp}{\bf Influence of the domain size in the vertical direction.} Relative error is calculated for the aerodynamic force magnitude and the aerodynamic power,
and plotted versus the distance from the `ceiling' of the flow domain.}
\end{center}
\end{figure}

Figure~\ref{fig:domain_size_cmp} shows the relative error for the
aerodynamic force magnitude, $(F_{abs,ave,OGE,6R}-F_{abs,ave,OGE,8R})/F_{abs,ave,OGE,8R} \cdot 100\%$,
and for the aerodynamic power, $(P_{ave,OGE,6R}-P_{ave,OGE,8R})/P_{ave,OGE,6R} \cdot 100\%$.
The horizontal axis shows the distance from the insect body point of reference to the `ceiling',
\textit{i.e.}, the penalization layer at the top of the domain, normalized to the wing length $R$.
We use the data for the case $L_z=8R$ to calculate this distance.
When the domain size is equal to $L_z=6R$, the duration of the takeoff spans four complete wingbeats,
and the wings leave the domain during the 5th wingbeat.
Hence, the 4th wingbeat is the most sensitive to the aerodynamic interaction with domain boundary.
The maximum difference for the force and for the power is 0.6\% and 1.6\%, respectively.
During the first three wingbeats, the error is less than 0.5\%.

\section*{Acknowledgments}
Numerical simulations were carried out using HPC resources of IDRIS, Paris (project 81664) and of Aix-Marseille Universit\'e (project Equip@Meso).
DK gratefully acknowledges the financial support from the CRM-ISM Postdoctoral Fellowship
and from the JSPS Postdoctoral Fellowship.
TE and KS thank the Deutsch-Franz\"osische Hochschule / Universit\'e Franco-Allemande (DFH-UFA) for financial support.


%
%
%

\bibliographystyle{plain}
\bibliography{ground_effect}

\begin{thebibliography}{10}

\bibitem{Angot_etal_1999}
P.~Angot, C.-H. Bruneau, and P.~Fabrie.
\newblock A penalisation method to take into account obstacles in viscous
  flows.
\newblock {\em Numerische Mathematik}, 81:497--520, 1999.

\bibitem{Aono_etal_2008}
H.~Aono, F.~Liang, and H.~Liu.
\newblock Near- and far-field aerodynamics in insect hovering flight: an
  integrated computational study.
\newblock {\em Journal of Experimental Biology}, 211(2):239--257, 2008.

\bibitem{Asselin_1997}
M.~Asselin.
\newblock {\em An introduction to aircraft performance}.
\newblock AIAA education series. American Institute of Aeronautics \&
  Astronautics, 1997.

\bibitem{Bimbard_etal_2013}
G.~Bimbard, D.~Kolomenskiy, O.~Bouteleux, J.~Casas, and R.~Godoy-Diana.
\newblock Force balance in the take-off of a pierid butterfly: relative
  importance and timing of leg impulsion and aerodynamic forces.
\newblock {\em Journal of Experimental Biology}, 216(18):3551--3563, 2013.

\bibitem{Card_Dickinson_2008}
G.~Card and M.~Dickinson.
\newblock Performance trade-offs in the flight initiation of {D}rosophila.
\newblock {\em Journal of Experimental Biology}, 211(3):341--353, 2008.

\bibitem{Chen_Sun_2014}
M.~W. Chen and M.~Sun.
\newblock Wing/body kinematics measurement and force and moment analyses of the
  takeoff flight of fruitflies.
\newblock {\em Acta Mechanica Sinica}, 30(4):495--506, 2014.

\bibitem{Chen_etal_2013}
M.~W. Chen, Y.~L. Zhang, and M.~Sun.
\newblock Wing and body motion and aerodynamic and leg forces during take-off
  in droneflies.
\newblock {\em Journal of The Royal Society Interface}, 10(89), 2013.

\bibitem{Chirarattananon_etal_2013}
P.~Chirarattananon, K.~Y. Ma, and R.~J. Wood.
\newblock Adaptive control for takeoff, hovering, and landing of a robotic fly.
\newblock In {\em IEEE International Conference on Intelligent Robots and
  Systems}, pages 3808--3815, 2013.

\bibitem{Dickinson_etal_1999}
M.~H. Dickinson, F.-O. Lehmann, and S.~P. Sane.
\newblock Wing rotation and the aerodynamic basis of insect flight.
\newblock {\em Science}, 284:1954--1960, 1999.

\bibitem{Engels_etal_2015}
T.~Engels, D.~Kolomenskiy, K.~Schneider, and J.~Sesterhenn.
\newblock {FluSI}: A novel parallel simulation tool for flapping insect flight
  using a {F}ourier method with volume penalization.
\newblock {\em arXiv:1506.06513}, 2015.

\bibitem{Engels_etal_2014}
T.~Engels, D.~Kolomenskiy, K.~Schneider, and J.~Sesterhenn.
\newblock Numerical simulation of fluid-structure interaction with the volume
  penalization method.
\newblock {\em Journal of Computational Physics}, 281:96--115, 2015.

\bibitem{Fontaine_etal_2009}
E.~I. Fontaine, F.~Zabala, M.~H. Dickinson, and J.~W. Burdick.
\newblock Wing and body motion during flight initiation in {D}rosophila
  revealed by automated visual tracking.
\newblock {\em Journal of Experimental Biology}, 212(9):1307--1323, 2009.

\bibitem{Gao_Lu_2008}
T.~Gao and X.~Lu.
\newblock Insect normal hovering flight in ground effect.
\newblock {\em Physics of Fluids}, 20:087101, 2008.

\bibitem{Ishihara_etal_2009}
D.~Ishihara, T.~Horie, and M.~Denda.
\newblock A two-dimensional computational study on the fluid–structure
  interaction cause of wing pitch changes in dipteran flapping flight.
\newblock {\em Journal of Experimental Biology}, 212(1):1--10, 2009.

\bibitem{Ishihara_etal_2014}
D.~Ishihara, T.~Horie, and T.~Niho.
\newblock An experimental and three-dimensional computational study on the
  aerodynamic contribution to the passive pitching motion of flapping wings in
  hovering flies.
\newblock {\em Bioinspiration \& Biomimetics}, 9(4):046009, 2014.

\bibitem{Jones_PhD_2005}
B.~L. Jones.
\newblock {\em Experimental investigation into the aerodynamic ground effect of
  a tailless chevron-shaped {UCAV}}.
\newblock Ph.d. thesis, Department of the Air Force, Air University, Air Force
  Institute of Technology, Wright-Patterson Air Force Base, Ohio, USA, 2005.

\bibitem{Kim_etal_2015}
E.~J. Kim, M.~Wolf, V.~M. Ortega-Jimenez, S.~H. Cheng, and R.~Dudley.
\newblock Hovering performance of {A}nna{\textquoteright}s hummingbirds
  ({C}alypte anna) in ground effect.
\newblock {\em Journal of The Royal Society Interface}, 11(98):20140505, 2014.

\bibitem{Kolomenskiy_etal_11b}
D.~Kolomenskiy, H.~K. Moffatt, M.~Farge, and K.~Schneider.
\newblock Two- and three-dimensional numerical simulations of the
  clap--fling--sweep of hovering insects.
\newblock {\em Journal of Fluids and Structures}, 27(5-6):784--791, 2011.

\bibitem{Kolomenskiy_Schneider_2009}
D.~Kolomenskiy and K.~Schneider.
\newblock A {F}ourier spectral method for the {N}avier--{S}tokes equations with
  volume penalization for moving solid obstacles.
\newblock {\em Journal of Computational Physics}, 228:5687--5709, 2009.

\bibitem{Leishman_2006}
J.~G. Leishman.
\newblock {\em Principles of helicopter aerodynamics}.
\newblock Cambridge University Press, 2006.

\bibitem{Liu_etal_2009}
Y.~Liu, N.~Liu, and X.~Lu.
\newblock Numerical study of two-winged insect hovering flight.
\newblock {\em Advances in Applied Mathematics and Mechanics}, 1(4):481--509,
  2009.

\bibitem{Maeda_PhD_2014}
M.~Maeda.
\newblock {\em Aerodynamics of flapping flight interacting with environments}.
\newblock Ph.d. thesis, Biomechanical Engineering Laboratory, Graduate School
  of Engineering, Chiba University, 2014.

\bibitem{Maeda_Liu_2013}
M.~Maeda and H.~Liu.
\newblock Ground effect in fruit fly hovering: A three-dimensional
  computational study.
\newblock {\em Journal of Biomechanical Science and Engineering},
  8(4):344--355, 2013.

\bibitem{Quinn_etal_2014}
D.~B. Quinn, K.~W. Moored, P.~A. Dewey, and A.~J. Smits.
\newblock Unsteady propulsion near a solid boundary.
\newblock {\em Journal of Fluid Mechanics}, 742:152--170, 3 2014.

\bibitem{Rayner_1991}
J.~M.~V. Rayner.
\newblock On the aerodynamics of animal flight in ground effect.
\newblock {\em Philosophical Transactions of the Royal Society of London.
  Series B: Biological Sciences}, 334(1269):119--128, 1991.

\bibitem{Schneider_2005}
K.~Schneider.
\newblock Numerical simulation of the transient flow behaviour in chemical
  reactors using a penalisation method.
\newblock {\em Computers \& Fluids}, 34:1223--1238, 2005.

\bibitem{Tanaka_etal_2011}
H.~Tanaka, J.~P. Whitney, and R.~J. Wood.
\newblock Effect of flexural and torsional wing flexibility on lift generation
  in hoverfly flight.
\newblock {\em Integrative and Comparative Biology}, 51(1):142--150, 2011.

\bibitem{Tanida_2001}
Y.~Tanida.
\newblock Ground effect in flight.
\newblock {\em JSME International Journal Series B}, 44(4):481--486, 2001.

\bibitem{Tobalske_etal_2007}
B.~W. Tobalske, D.~R. Warrick, C.~J. Clark, D.~R. Powers, T.~L. Hedrick, G.~A.
  Hyder, and A.~A. Biewener.
\newblock Three-dimensional kinematics of hummingbird flight.
\newblock {\em Journal of Experimental Biology}, 210(13):2368--2382, 2007.

\bibitem{Truong_etal_2013}
T.~V. Truong, D.~Byun, M.~J. Kim, K.~J. Yoon, and H.~C. Park.
\newblock Aerodynamic forces and flow structures of the leading edge vortex on
  a flapping wing considering ground effect.
\newblock {\em Bioinspiration \& Biomimetics}, 8(3):036007, 2013.

\bibitem{Wootton_etal_2003}
R.~J. Wootton, R.~C. Herbert, P.~G. Young, and K.~E. Evans.
\newblock Approaches to the structural modelling of insect wings.
\newblock {\em Philosophical Transactions of the Royal Society of London B:
  Biological Sciences}, 358(1437):1577--1587, 2003.

\bibitem{Wu_etal_2014}
J.~Wu, C.~Shu, N.~Zhao, and W.~Yan.
\newblock Fluid dynamics of flapping insect wing in ground effect.
\newblock {\em Journal of Bionic Engineering}, 11:52--60, 2014.

\bibitem{Zumstein_etal_2004}
N.~Zumstein, O.~Forman, U.~Nongthomba, J.~C. Sparrow, and C.~J.~H. Elliott.
\newblock Distance and force production during jumping in wild-type and mutant
  {D}rosophila melanogaster.
\newblock {\em Journal of Experimental Biology}, 207(20):3515--3522, 2004.

\end{thebibliography}




\end{document}